\documentstyle[psfig,amssymb,graphicx,epsf]{mn}

\title[Optical and Near-Infrared Observations of Faint Radio  Sources] 
{The Phoenix Survey: Optical and Near-Infrared Observations of Faint Radio
  Sources}

\author[A.Georgakakis {\it et al.}] {A. Georgakakis$^{1,2}$, B.
  Mobasher$^{2}$, L. Cram$^{3}$, A. Hopkins$^{3}$, C. Lidman$^{4}$,\\ \\ 
  {\LARGE M. Rowan-Robinson$^{2}$} \\ \\ 
  $^1$ School of Physics and Astronomy, University of Birmingham,
  Edgbaston, B15 2TT, UK\\
  $^2$Astrophysics Group, Blackett Laboratory, Imperial College, Prince
  Consort Rd , London SW7 2BZ, UK\\
  $^3$ Astrophysics Department, School of Physics, University of Sydney,
  NSW, Australia 2006\\ 
  $^4$ European Southern Observatory, Casilla 19001, Santiago 19, Chile}

\begin{document}
\maketitle  

\begin{abstract}
  Using a deep Australia Telescope Compact Array (ATCA) radio survey
  covering an area of $\approx$3\,deg$^{2}$ to a 4$\sigma$ sensitivity of 
  $\ge100$\,$\mu$Jy at 1.4 GHz, we study the nature of faint radio galaxies. 
  About $50\%$ of the detected radio sources are identified with an optical 
  counterpart revealed by CCD photometry to m$_{R}$=22.5\,mag. Near-infrared 
  ($K$-band) data are also available for a selected sample of the radio 
  sources, while spectroscopic
  observations have been carried out for about $40\%$ of the optically
  identified sample.  These provide redshifts and information on the
  stellar content. Emission-line ratios imply that most of the emission
  line sources are star-forming galaxies, with a small contribution
  ($\approx10\%$) from Sy1/Sy2 type objects.  We also find a significant
  number of absorption line systems, likely to be ellipticals. These
  dominate at high flux densities ($>$1\,mJy) but are also found at sub-mJy
  levels.  Using the Balmer decrement we find a visual extinction
  $A_{V}$=1.0 for the star-forming faint radio sources.  This
  moderate reddening is consistent with the $V-R$ and $R-K$ colours of the
  optically identified sources. For emission line galaxies, there is a
  correlation between the radio power and the H$\alpha$ luminosity, in
  agreement with the result of Benn {\it et al.} (1993). This suggests that
  the radio emission of starburst radio galaxies is a good indicator of
  star-formation activity.
\end{abstract} 
 

\section{Introduction}
Deep radio surveys (Windhorst {\it et al.} 1985; Windhorst {\it et al.}
1993; Fomalont {\it et al.} 1997; Hopkins {\it et al.} 1998) have revealed
a flattening in the normalized source counts below S$_{1.4} \approx
$10\,mJy suggesting the emergence of a new population of faint radio
sources.  Photometric (Thuan {\it et al.} 1984, 1992; Kron {\it et al.}
1985; Windhorst {\it et al.} 1985; Thuan \& Condon 1987) and spectroscopic
studies (Benn {\it et al.} 1993) reveal that 
the sources responsible for the flattening are predominantly star-forming 
galaxies, similar to those of the IRAS starburst population. The proportion
of active galactic nuclei (AGNs) is much smaller than it is at higher flux
densities (Benn {\it et al.} 1993; Gruppioni {\it et al.} 1998; Kron 
{\it et al.} 1985).  However, the local density of star-forming radio 
sources is too small to explain the turn-up point and slope of the 
$\log N / \log S$ curve.  Moreover, evolutionary models of the radio sources 
that dominate the counts at high flux densities cannot reproduce these 
features either  (Danese {\it et al.} 1987; Danese, De Zotti \& 
Franceschini 1985).  Therefore, models for the source counts invoke strong 
evolution of either spiral galaxies (Condon 1989; Dunlop \& Peacock 1990), or
star-forming IRAS population galaxies (Danese {\it et al.} 1987;
Rowan-Robinson {\it et al.} 1993; Hopkins {\it et al.} 1998). 

Visual inspection of faint radio sources shows that many have optical
counterparts which are preferentially located in pairs or small groups
exhibiting disturbed optical morphologies, suggestive of interactions or
mergers (Kron {\it et al.} 1985; Gruppioni {\it et al.} 1998).  Although
this method is not an ideal way to identify physically associated groups,
the frequency of cases suggests that such phenomena are partially responsible
for the enhanced star formation rate seen in these objects (Kron {\it et
  al.} 1985; Windhorst {\it et al.} 1995).

Further evidence for the appearance of a new population at faint radio flux
densities comes from measurements of angular sizes. These show that at flux
densities below a few mJy, the median angular size drops abruptly to few
arcseconds (Coleman \& Condon 1985; Oort {\it et al.} 1987). This 
can be reconciled if faint radio sources are associated with spiral
galaxies with radio morphology more compact than that of ellipticals 
which dominate at higher flux densities (Coleman \& Condon 1985).

Despite the effort devoted to the study of the faint radio population there
is still no clear answer to the following questions: (i) what is the
stellar population in sub-mJy radio sources and how do they evolve with
redshift, (ii) what is their relation to the local population of normal 
galaxies and (iii) how are the star-formation activity and radio properties 
of these objects related. To address these points  we present 
a deep and homogeneous radio  (1.4\,GHz) survey (Phoenix), covering 
an area of 3\,deg$^{2}$ combined with optical ($V$, $R$-band) and near 
infrared ($K$-band) photometry and optical spectroscopy.

The observations, astrometric calibration, catalogue generation and optical
identifications are described in section 2. The final catalogue is
presented in section 3, while the data analysis and the results are
discussed in section 4. Finally, section 5 summarises our conclusions.
Throughout this paper we assume a value $H_{0}=50$\,km\,s$^{-1}$\,Mpc$^{-1}$
and $q_{0}=0.5$.

\section{Observations}

\subsection{Radio Observations} 

\begin{figure}
\centerline{\psfig{figure=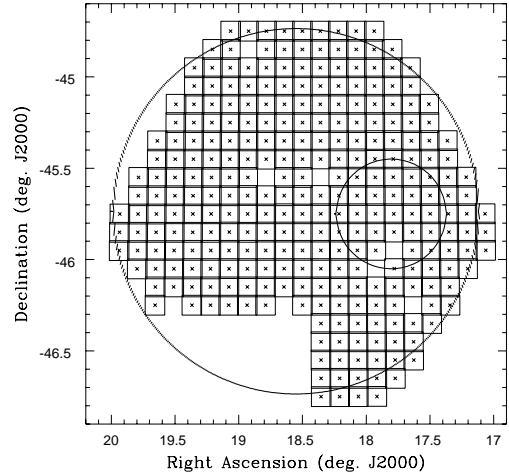,width=0.45\textwidth,angle=0}} 
\caption{The area covered by the Phoenix Deep Field (large disk)
  and the Phoenix Deep Field Subregion (small disk). Individual CCD
  $R$-band frames are depicted as squares with a cross in the middle}\label{PDF} 
\end{figure}

The radio observations were made at 1.46\,GHz using the 6A configuration of
the Australia Telescope Compact Array (ATCA).  The mosaic of 30 pointing
centres covers a $2^{\circ}$ diameter area centered at ${\mathrm
  RA}(2000)=01^{\mathrm h}~14^{\mathrm m}~12\fs16$; ${\mathrm
  Dec.}(2000)=-45^{\circ}~44'~8\farcs0$ (galactic latitude $b=-71^{\circ}$). 
This field, hereafter referred to as the Phoenix Deep Field (PDF), is 
shown as the large disk in Figure \ref{PDF}. Additional observations were 
made with a single pointing centered at
  ${\mathrm RA}(2000)=01^{\mathrm h}~11^{\mathrm m}~13\fs0$; ${\mathrm
  Dec.}(2000)=-45^{\circ}~45'~0\farcs0$. This is referred to as the Phoenix
Deep Field Sub-region (PDFS), covering 36\,arcmin diameter and shown in
Figure \ref{PDF} as the small disk.  The synthesised beam FWHM for the 
PDFS and for each of the PDF pointing centres is $\approx$6 and 
$\approx$8\,arcsec respectively.

Details of the observations, image formation, source extraction and
catalogue generation are presented in Hopkins {\it et al.} (1998).  In
summary, a source is included into the catalogue if its peak flux density
is 4$\sigma$ above the local RMS and if it is confirmed by visual
inspection. A total of 938 sources with fluxes S$_{1.4}>0.2$\,mJy are 
detected within the PDF and 232 sources are extracted from the 36\,arcmin 
diameter area of the PDFS, to the limiting flux of 0.1\,mJy. The combined 
radio catalogue consists of 1079 individual sources identified on both fields. 
There are two kinds of incompleteness in the catalogue, as
with any sample limited by peak flux density. The first is a loss of
sensitivity due to the attenuation of the primary beam away from a field
center. This has been minimized in the Phoenix survey by the mosaicing
strategy used. The second is the fact that extended objects with a total
flux density above the survey limit can be missed by an algorithm which
initially detects candidates based on their peak flux densities. Methods of
correcting these effects have been described by Hopkins {\it et al.}
(1998). The PDF catalogue is $\approx80\%$ complete to 0.4\,mJy, while
the PDFS is $\approx90\%$ complete to 0.15\,mJy.

\subsection{Optical Photometric Observations}

\subsubsection{Observations and data reduction}

Optical CCD photometry of the Phoenix Field was carried out at the
Anglo-Australian Telescope (AAT) during two observing runs in 1994, 1995
September, in the $V$ and $R$-bands.  In both runs, the TEK 1K CCD camera
was mounted at the $f$/3.3 prime focus, giving an image scale of
0.39$^{\prime \prime}$pixel$^{-1}$ and a field of view of $6.5^{\prime}
\times6.5^{\prime}$. The exposure time was 120 and 60\,sec for the $V$ and
$R$-band observations respectively.  Between successive pointings the
telescope was offset by 6$^{\prime}$ in right ascension or declination.
About $85\%$ areal coverage was achieved in the $R$-band (Fig. \ref{PDF}).  
However, due to poor weather conditions, only half of the Phoenix field 
has yet been surveyed in the $V$-band.

The optical data were reduced following standard procedures, using IRAF
tasks. Firstly, the optical CCD frames were debiased using the overscan
strip. Secondly, dome-flats were employed to flat-field the images.  Bad
pixel columns were removed by interpolating between adjacent columns.
Photometric calibration was performed using standard stars from Landolt
(1992). These provide both the instrumental zero point and the atmospheric 
extinction relation due to airmass. The photometric accuracy, estimated 
using these standard stars, is better than $\approx$0.01\,mag for both $V$ 
and $R$-band observations.
 
\subsubsection{Catalogue Construction}

The sources in the reduced optical frames were extracted using the FOCAS
package (Jarvis \& Tyson 1981). The main input parameters are the
detection threshold, given as a multiple of the sky variance
($\sigma_{Sky}$) and the minimum area in pixels for an object to be
extracted. After considerable experimentation, we adopted a threshold of
2.5$\times \sigma_{Sky}$ and a minimum area of 25 pixels
($\approx$4\,arcsec$^{2}$).  This choice of values minimises the number of
spurious detections, while ensuring that most faint galaxies are
successfully extracted.  For each individual frame we masked out vignetted
corners and regions contaminated by bright stars, to avoid obviously
spurious detections.  The catalogues generated for each individual frame
were then merged into a final catalogue. Because of the overlap between
adjacent frames, a number of objects are recorded more than once. These
records are replaced by a single object in the catalogue. A final visual
inspection removed any remaining spurious detections. The completeness
limit of the $R$-band catalogue is found to be $\approx22.5$\,mag 
(Appendix $A$).

\subsubsection{Astrometry}

The astrometry on the optical images is carried out using the APM catalogue
to locate $\approx$10 stars on each frame. Using the STARLINK ASTROM
software, an astrometric model relating the pixel positions of those stars
to their astrometric positions was calculated for each frame. The
astrometric coordinates of all the objects on that frame were then
determined. To estimate the accuracy of the astrometry, the positions of
overlapping sources on the APM plates and on our CCD frames were compared.
They were found to have an RMS of $\approx0.7^{\prime \prime}$.

\subsubsection{Optical identification of the radio sources}

To identify optically the sources detected in the radio survey we used the
$R$-band observations, which are deeper than the $V$-band data and have the 
more complete areal coverage.  We followed the technique employed by 
Downes {\it et al.}
(1986), calculating the probability that a given candidate is the
identification by using Bayes' theorem: Consider an optically detected
candidate with magnitude $m$ at a distance $r$ from the radio position.
Given the surface density of objects brighter than $m$, $\Sigma(<m)$, the
expected number of candidates within $r$ is

\begin{equation}
\mu=\pi\,r^{2}\,\Sigma(<m).
\end{equation}

\noindent Assuming that source positions are Poissonian, the probability 
of at least one object brighter than $m$ within radius $r$ is

\begin{equation}
P=1-exp(-\mu),
\end{equation}

\noindent which reduces to $\mu$ for $\mu<<$1. In this case, the candidate 
is unlikely to be a chance association. We apply an upper limit in the search
radius, $r<10^{\prime\prime}$ and a cutoff in the probability, $P<0.05$, to 
limit the optical identifications to those candidates that are least likely 
to be spurious alignments (Hopkins 1998a).  The background density of the 
objects (both galaxies and stars) is calculated from our $R$-band observations 
to the limiting magnitude of $m_{R}=22.0$\,mag (Appendix $A$).  At fainter
magnitudes we used the surface density of objects found by Metcalfe {\it et
al.} (1995). As a result, we propose 504 identifications to the limiting
magnitude m$_{R}=22.5$\,mag for a total of 1079 radio sources
($\approx47\%$). 

\subsubsection{Optical Photometry}

To estimate colours, the flux in the $V$ and $R$-bands is integrated within
an effective  aperture corresponding to a physical size of $\approx$15\,kpc 
radius at the
redshift of the galaxy.  This choice of aperture ensures that a substantial
fraction of the galaxy light is enclosed. This minimizes the effect of
colour gradients while increasing the S/N, especially at faint magnitudes
where the uncertainties are dominated by the photon statistics of the sky
background.

To facilitate comparison with other studies and to recognize that the
aperture scheme used to estimate colour will tend to underestimate the
fluxes of the brightest and most extended sources we used the `total'
magnitude calculated by FOCAS  as an estimate of the total galaxy optical 
flux. The FOCAS `total' magnitude is calculated
by integrating the luminosity within an aperture defined by adding rings 
around the object until the detection area is exceeded by a factor of 2.


\subsection{Near-Infrared imaging}

The near-infrared observations used the IRAC2b camera at the 2.2m European
Southern Observatory (ESO) telescope at La Silla and CASPIR at the 2.3m
Australian National University (ANU) telescope.  Because of the small
detector format of the infrared arrays, pointed observations of a total
of 46 individual radio sources were carried out.

\begin{enumerate}
\item The ESO observations were carried out on 1991 September 24-27 at the
  Cassegrain focus of the 2.2m telescope.  IRAC2b is a 256$\times$256 pixel
  NIMCOS-3 detector with 0.5$^{\prime \prime}$\,pixel$^{-1}$, giving a field of
  view of 2$^{\prime}\times$ 2$^{\prime}$.  We observed 34 radio sources
  within the PDFS, having sub-mJy radio flux densities and optical magnitudes
  in the range $19.0<m_{R}<22.0$.

  Dithering observations of each source consisted of a sequence of 6
  integrations of 10\,sec each, or of 4 integrations of 15\,sec each, 
  followed by
  an offset of the telescope.  The number of integrations and the exposure
  time were adjusted to the seeing conditions, to avoid saturation of the
  array.  To get accurate sky frames, the offset vector was changed between
  successive observations. The total exposure time varied from 15\,min
  to 1\,h depending on the optical magnitude of the candidate. The filter 
  used was $K^{\prime}$ (Wainscoat and Cowie 1992).

  After dark-field subtraction, each image was flat-fielded using dome
  flats. An illumination correction, accounting for the large scale 
  sensitivity gradients of the array, was also applied
  (see ESO manual\footnote{http://epu.ls.eso.org/lasilla/Telescopes/2p2T/E2p2M/IRAC2}).
  The sky frame to be subtracted from a given target frame was generated by 
  median combining the 20 images closest in time to the frame in question. 
  The sky-subtracted frame was then smoothed using a $100 \times 100$ pixel
  median filter, designed to preserve as much as possible the photometric
  integrity of the data. All the images of the same source were coadded to
  produce the final image.

  Photometric calibration used standards from Carter and Meadows (1995).
  The zero points in the calibration equations were stable, with a
  night-to-night variation less than 0.10\,mag.  The uncertainty in the
  zero point is estimated to be less than $\approx 0.02$.

\item The ANU near-infrared observations were carried out on 1997 November
  8-10 at the Cassegrain focus of the 2.3m telescope.  CASPIR uses a Santa
  Barbara Research Center CRC463 $256 \times 256$ InSb detector array with a
  scale of 0.5$^{\prime \prime}$\,pixel$^{-1}$, giving a field of view of
  2$^{\prime}\times$ 2$^{\prime}$.  A total of 12 sub-mJy radio sources
  with optical counterparts brighter than $m_{R}=19.0$\,mag were observed.
  These observations were carried out in non-photometric conditions.
 
  As with the ESO observations, a dithering technique was adopted,
  consisting of a sequence of 12 integrations of 5\,sec each followed by an
  offset of the telescope.  The total exposure time for each
  target object was 17\,min. The filter used was
  $K_{n}$\footnote{http://msowww.anu.edu.au/observing/docs/manual} which is somewhat 
  different from $K^{\prime}$. However, the difference between the two 
  bands is relatively small ($\le0.1$\,mag) and does not affect any of our 
  conclusions.

  The data reduction followed a procedure similar to that described for the
  ESO observation.  Photometric calibration was performed using faint
  standard stars from Carter and Meadows (1995).  A mean extinction
  relation was derived for all the nights together. Because of the
  non-photometric conditions, the uncertainty in the estimated zero point
  in the calibration equation was $\approx$0.14\,mag.  However, both the zero
  point and the extinction coefficient agree, within the errors, with
  typical values reported in the CASPIR manual (ZP$_{Kn}$=20.5,
  extinction=0.10\,mag\,airmass$^{-1}$).
\end{enumerate}


\subsection{Spectroscopic Observations}

Spectroscopic data were obtained using slit spectroscopy at the ESO 3.6\,m
telescope and multi-object fibre spectroscopy at the 2 degree field
spectroscopic facility (2dF) at the AAT.

\subsubsection{Slit Spectroscopy}

The spectra of 20 of the 34 candidates observed in K$^{\prime}$ were
obtained using EFOSCI at the Cassegrain focus of the 3.6m ESO telescope at
La Silla, on 1996 October 9-10.  Two gratings were used, covering the
wavelength ranges 5940-9770\,\AA\, (7.5\,\AA\,pixel$^{-1}$) and 
3640-6860\,\AA\,
(6.3\,\AA\,pixel$^{-1}$). The spatial scale of the detector is 0.61$^{\prime
\prime}$\,pixel$^{-1}$.  A 1.5$^{\prime\prime}$ slit on the first night and
a 2.0$^{\prime\prime}$ slit during the second night were used, with an
integration time varying from 10 to 30\,min depending on the brightness of
the object. 

Data reduction was carried out using the Starlink FIGARO package. Firstly,
the bias frames were co-added and then subtracted from the target frames.
Secondly, dome-flats, produced by a halogen lamp, were employed to
flat-field the data.  The wavelength calibration was carried out using a
helium-argon spectral lamp, while a relative flux calibration was performed
using LTT377 spectrophotometric standard (Hamuy {\it et al.} 1994).
Finally, a wavelength dependent correction for extinction was applied.
Values for the extinction parameter at La Silla were taken from the ESO
observers' manual.
        
Although the faintest spectra have low S/N, all but one of the target
objects were assigned a redshift based on one or more emission lines.

\subsubsection{2-degree Field Spectroscopy}

Multi-fibre spectroscopy was carried out using 2dF at the prime focus of
AAT. The 2dF instrument is designed to allow simultaneous acquisition of up
to 400 spectra of objects within a 2$^{\circ}$ diameter field on the sky. It
consists of two spectrographs, each having 200 fibres and two $1024 \times
1024$ thinned Tektronix CCDs each receiving 200 fibres.  The fibres are
$\simeq$2\,arcsec in diameter resulting in 2 pixel wide spectra on the
detectors.  The data for the present study were obtained during two
observing runs in 1996 December and 1997 September.  At the time, only one
of the two spectrographs was operating, resulting in a total of 200 spectra
per exposure.

The optical candidates selected for the 2dF observation had
m$_{R}<$21.5\,mag and were segregated into groups according to their
optical magnitude. The total exposure time varied from 1.5\,h for the
brightest objects to 3.5\,h for the optically fainter ones (carried out in
3 to 7 half-hour exposures).  The grating used was the 270R, having 75\%
throughput over the wavelength range 5000$<\lambda<$8500\,\AA, with a
dispersion of 199.9\,\AA\,mm$^{-1}$. The spectral resolution was 
$\lambda/ \Delta \lambda \simeq 600$ with a pixel scale of 
$\simeq 4.7$\,\AA\,pixel$^{-1}$.

The spectroscopic data reduction was performed using the 2DFDR package,
developed for the reduction of 2dF data.  No atmospheric extinction 
correction was applied to the spectra. This was found to be negligible, 
since most of the observations were carried out at relatively low zenith 
distances ($<40^{\circ}$). Redshifts 
were determined  by visual inspection of the resulting spectra. Each 
spectroscopically observed source
was assigned an index, $Q$, $0<Q<3$, indicating the quality of the redshift
measurement. A value $Q=3$ corresponds to three or more identified spectral
features, indicating a firmly established redshift. A value $Q=0$, 1 and 2
corresponds to none (no redshift determination), one and two identified
spectral features respectively. Redshifts were established for 221 out of
312 candidate optical identifications observed with 2dF. Since there was 
an overlap of 12 objects between the sources observed with ESO 3.6\,m
telescope and those observed using the 2dF, the total number of determined
redshifts, from both runs, is 228 out of 320.  The redshifts of the common
objects in the spectroscopic runs were found to be consistent with an 
RMS of $\approx0.001$. Furthermore, to measure the strengths of diagnostic
emission lines (see section 4.1), a relative flux calibration was performed
using a crude response function provided by the 2dF team.  This was
calculated by estimating the 2dF efficiency in 3 bands (B, V, R) and then
interpolating using a second order polynomial.

\begin{figure*}
\centering
\includegraphics[width=1.05\textwidth, height=1.05\textheight,angle=-180]{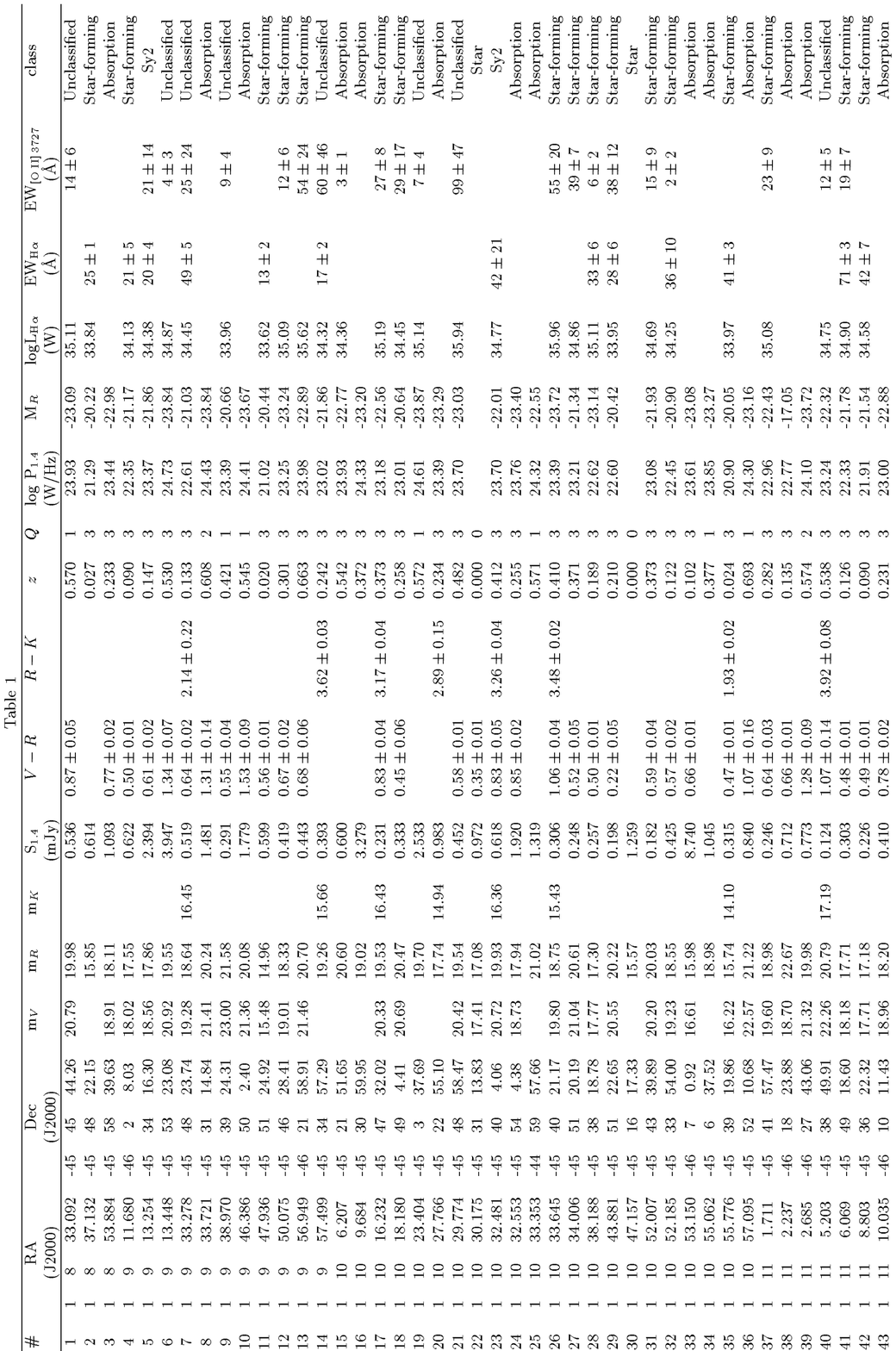}
\end{figure*}

\begin{figure*}
\centering
\includegraphics[width=1.05\textwidth, height=1.05\textheight,angle=-180]{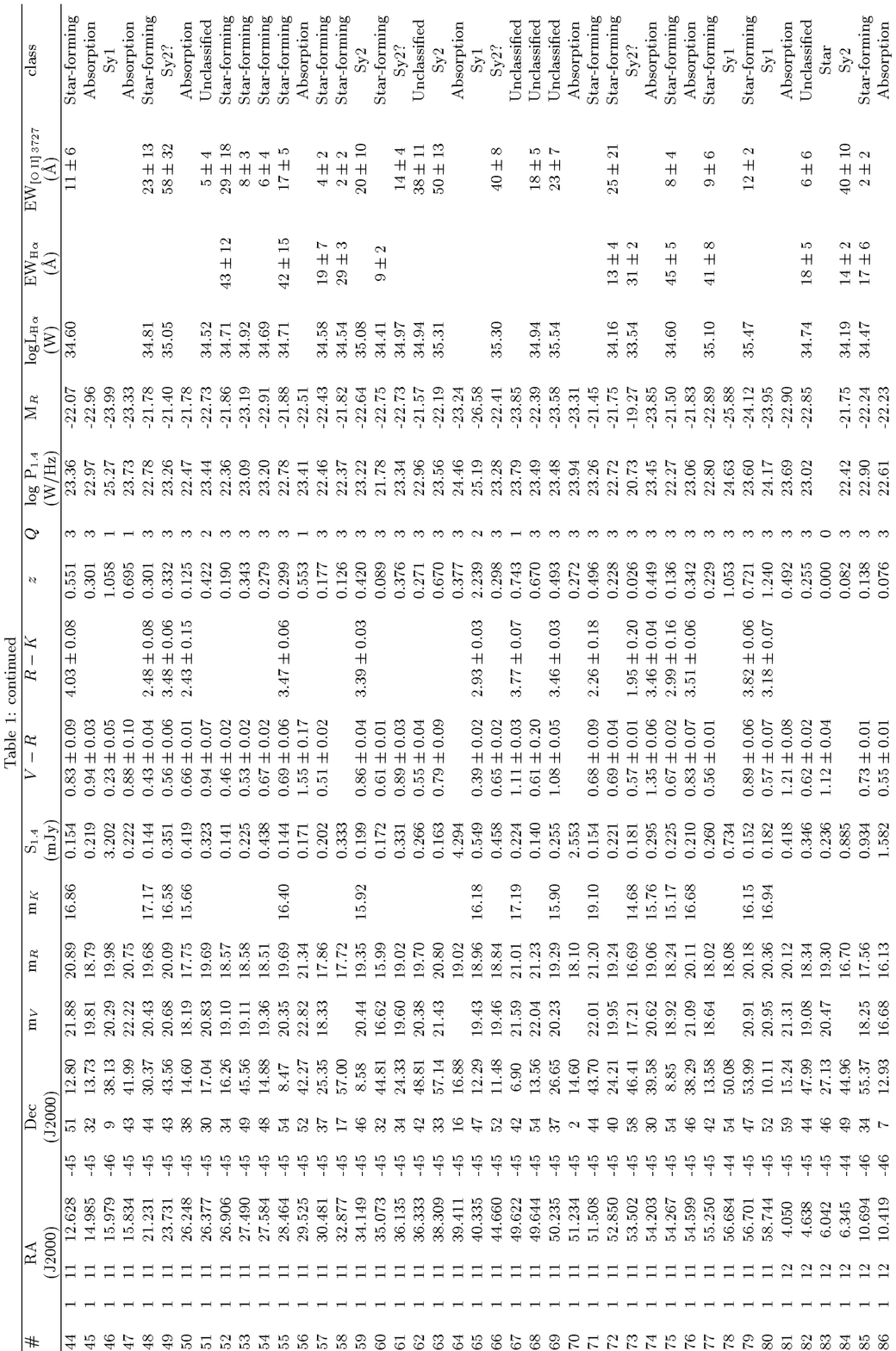}
\end{figure*}

\begin{figure*}
\centering
\includegraphics[width=1.05\textwidth, height=1.05\textheight,angle=-180]{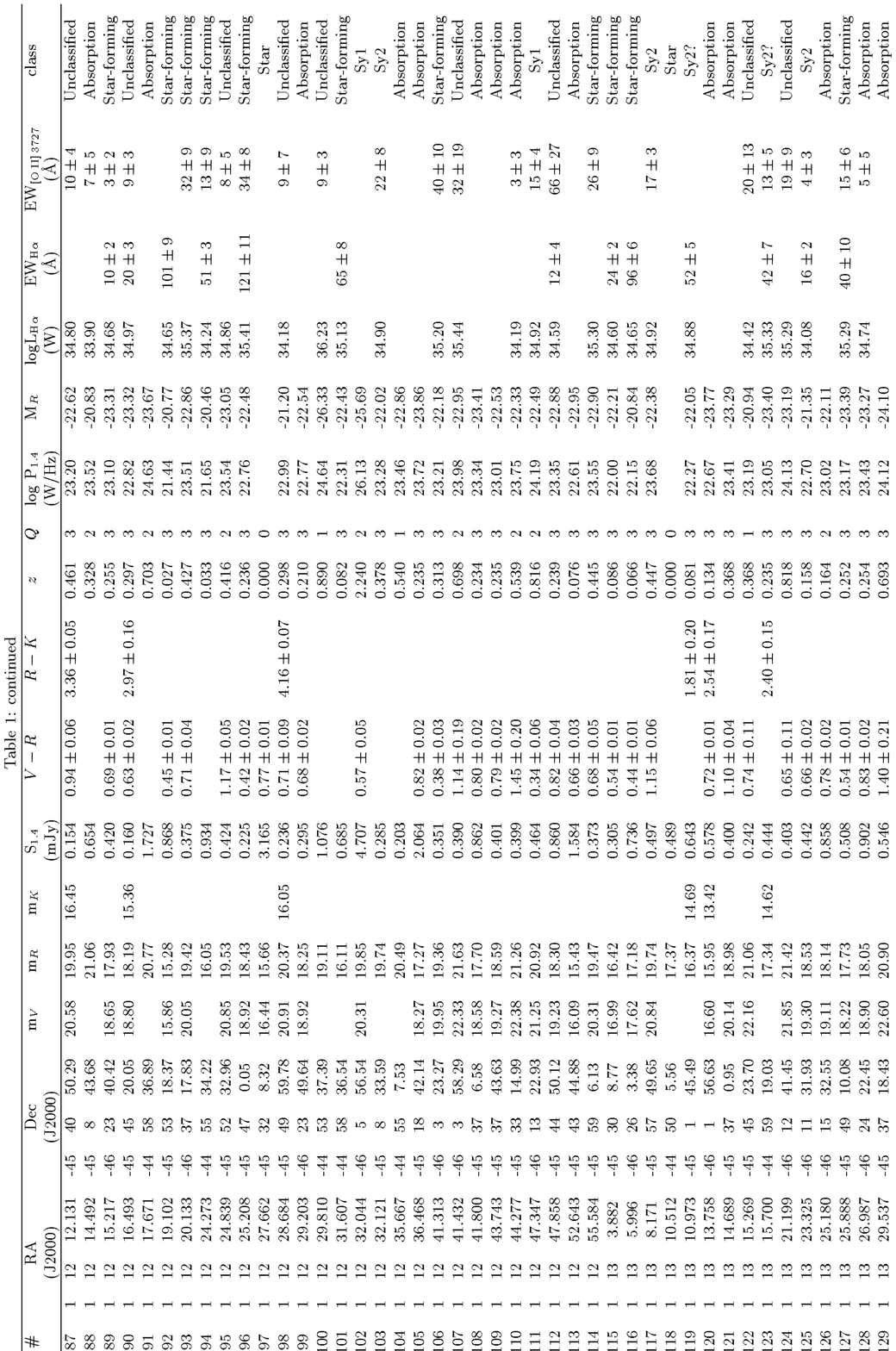}
\end{figure*}

\begin{figure*}
\centering
\includegraphics[width=1.05\textwidth, height=1.05\textheight,angle=-180]{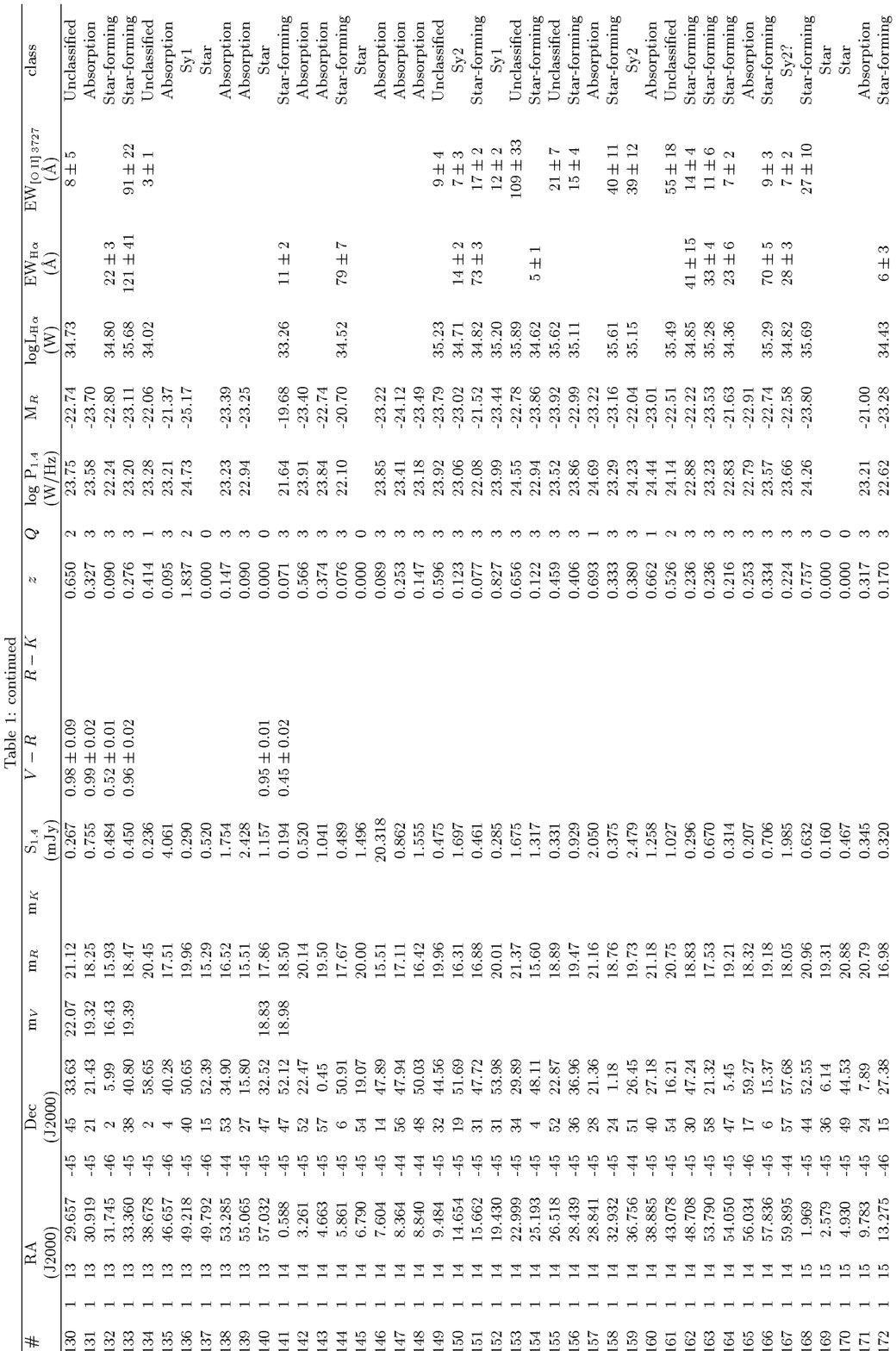}
\end{figure*}

\begin{figure*}
\centering
\includegraphics[width=1.05\textwidth, height=1.05\textheight,angle=-180]{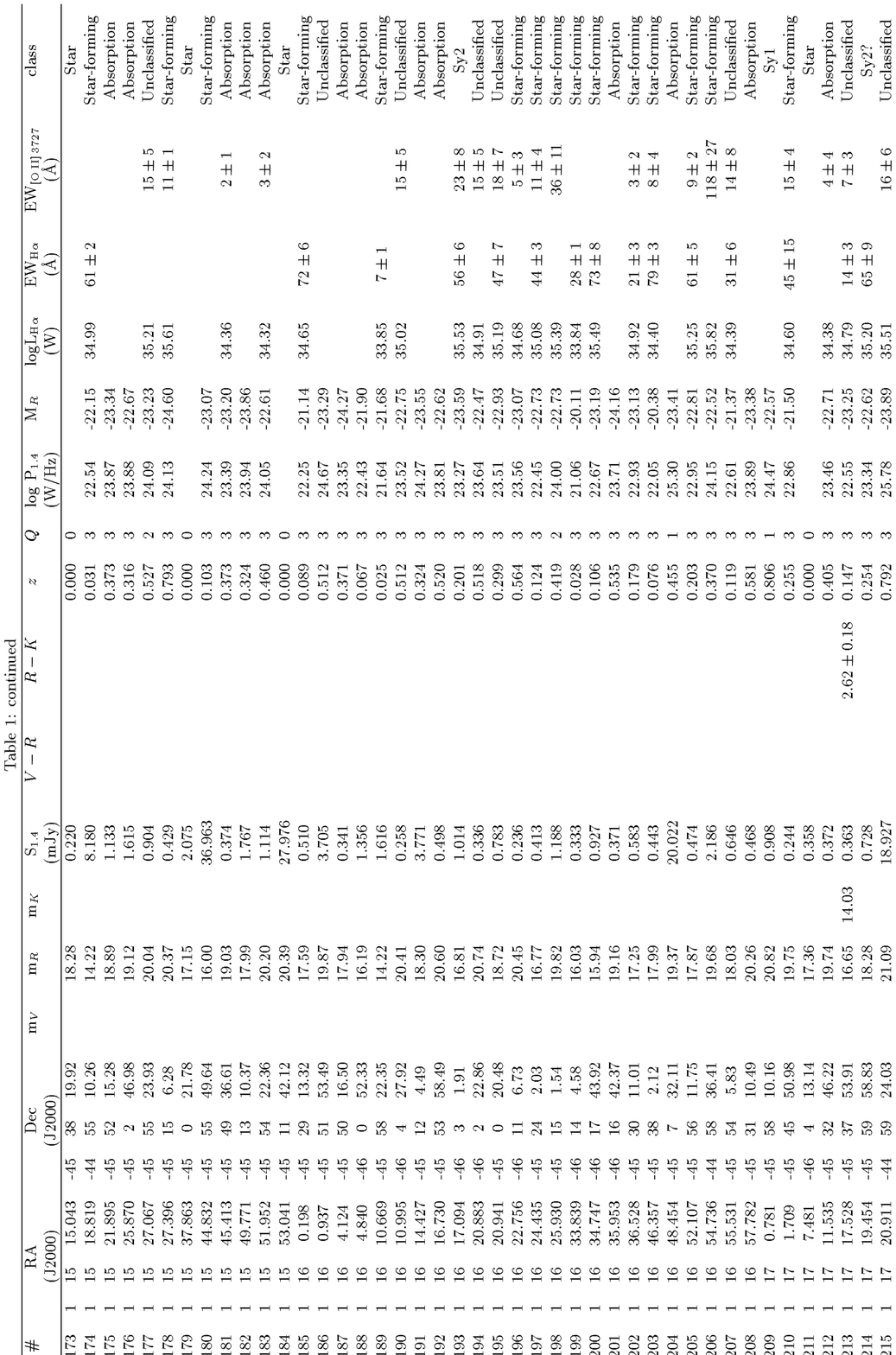}
\end{figure*}

\begin{figure*}
\centering
\includegraphics[width=1.05\textwidth, height=1.05\textheight,angle=-180]{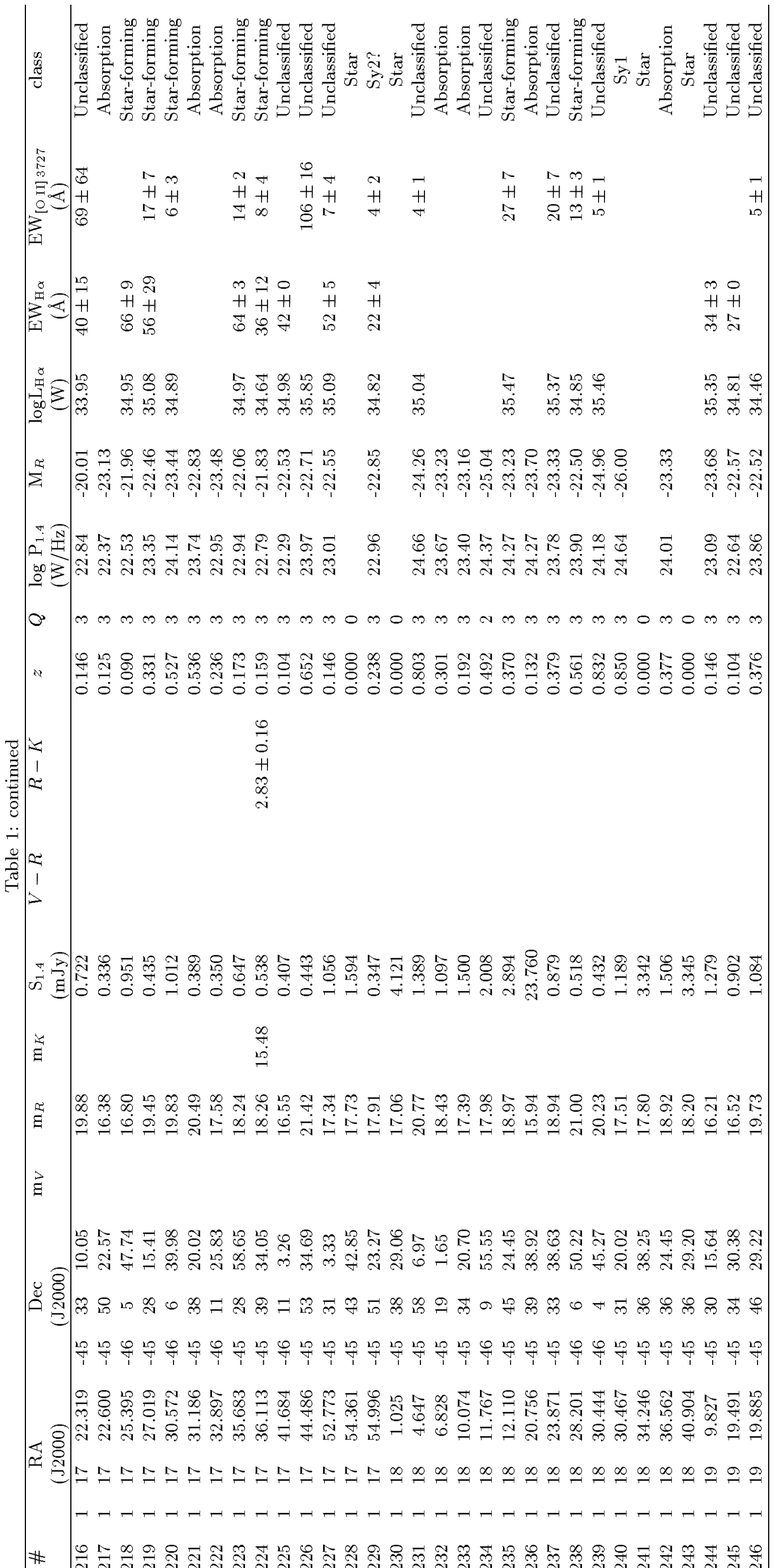}
\end{figure*}

\section{The Catalogue}
The photometric (optical and near-infrared), radio, spectroscopic and
intrinsic properties of the faint radio sources with established
redshift are presented in Table 1, which has the following format 

(1) Radio source sequential number.

(2-7) Right ascension and declination of the optical counterpart of the
radio source in J2000. 

(8-10) Apparent `total' magnitudes (section 2.2.5), m$_{V}$, m$_{R}$ and
m$_{K}$, of the optical counterpart of the radio source at $V$, $R$ and $K$
wavebands respectively.

(11) Radio flux density at 1.4\,GHz in mJy.

(12-13) $V-R$ and $R-K$ colours respectively, calculated using aperture
photometry (section 2.2.5) along with the associated uncertainties.  

(14) Redshift $z$.

(15) Redshift estimate quality parameter, $Q$, defined in section 2.4.2.

(16) $\log_{10}$ 1.4\,GHz radio luminosity in W\,Hz$^{-1}$.

(17) Absolute $R$-band magnitude, M$_{R}$, calculated using the `total'
$R$-band apparent magnitude, m$_{R}$.

(18) $\log_{10}$ H$\alpha$ luminosity in W.

(19) Rest-frame H$\alpha$ equivalent width in \AA.

(20) Rest-frame [O\,II]\,3727 equivalent width in \AA.

(21) Spectral classification, described in section 4.1.

\section{Results}

\subsection{Spectral Classification}

The optical counterparts of the radio sources in the sample are classified
using their optical spectral features. For this purpose, we used absorption
lines (Balmer lines, H+K\,4000\,\AA, G-band\,4220\,\AA, Mgb\,5175\,\AA,
NaD\,5893\,\AA, [Mg\,II]\,2796\,\AA), or the diagnostic emission line ratios
(Baldwin, Phillips \& Terlevich 1981; Veilleux \& Osterbrock 1987; Rola,
Terlevich \& Terlevich 1997). This latter analysis is hampered by the low
S/N of some of the spectra, the relatively poor sensitivity in the blue
(affecting [O\,II]\,3727, H$\beta$\,4861, [O\,III]\,5007) and by the
presence of strong atmospheric bands longward of 8000\,\AA, affecting
H$\alpha$ at $z>$0.25.  A further caveat is that stellar H$\beta$
absorption may reduce the measured flux and equivalent width of this line.
This is estimated to be $-1.7\pm0.3$\,\AA, for $z<0.3$ radio sources (Appendix
$B$).  This correction is only approximate and might not be applicable to
radio sources at $z>0.3$. Ideally, knowledge of the H$\beta$ absorption
in {\it individual} galaxies is needed for classification purposes.
Consequently, we decided not to correct for this effect and present the
classification based on measured fluxes. 

\begin{figure} 
\centerline{\psfig{figure=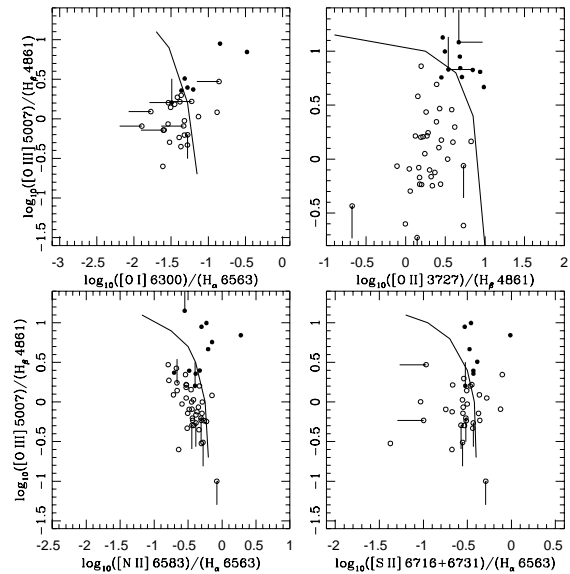,width=0.45\textwidth,angle=0}}
\caption{ The diagnostic emission line ratios used to classify the 
  optical counterparts of faint radio sources.  The solid curves, in all
  panels, separate Seyferts from galaxies heated by OB stars. For the 
  top right panel  the solid curve is taken from Rola, Terlevich \&
  Terlevich (1997), while for rest of the panels the solid curves are from
  Veilleux \& Osterbrock (1987). Filled circles denote objects classified
  as Seyfert 2 type objects and open circles signify star-forming
  galaxies. The bars denote upper or lower limits. 
  }\label{diagnostic}
\end{figure}

Diagnostic line ratios, consisting of H$\beta$/[O\,III] 5007, [S\,II]
6716+31/H$\alpha$, [O\,I] 6300/H$\alpha$ and [N\,II]
6583/H$\alpha$, were calculated. Figure \ref{diagnostic} shows how these 
ratios are
used to classify the galaxies. Equal weights were assigned to each of the
diagnostic line ratios available for a given object.  The class assigned to
the source was that consistent with the majority of the line ratios
available for that source.
Uncertainties in the flux calibration and dust extinction will not affect
these line ratios significantly, since the pairs lie close in wavelength.
However, the analysis can be applied only to objects with $z<0.3$. At
higher redshifts, H$\alpha$ moves out of the optical window and other
line ratios, [O\,II]\,3727/H$\beta$ and [O\,III]\,5007/ H$\beta$,
must be employed (Rola, Terlevich \& Terlevich 1997).  Figure  
\ref{diagnostic} shows how
these ratios allow classification of the central ionizing source.
However, the [O\,II]\,3727/H$\beta$ ratio is sensitive to both the
approximate flux calibration applied to the 2dF spectra and to extinction,
since it involves lines that are separated in wavelength.

The faint radio sources with redshift determination ($\approx70\%$ of the 
spectroscopic sample) were classified into four classes based on their
optical spectra: (i) galaxies with only absorption lines in their spectra
($\approx22\%$ of the spectroscopic sample); (ii) star-forming galaxies
($\approx25\%$); (iii) Seyfert galaxies ($\approx9\%$); and (iv)
unclassified objects ($\approx15\%$). The unclassified objects displayed
at least one identified emission line (allowing a redshift to be
determined), but the poor S/N, or a very small number of emission lines
within the observable window, or the presence of instrumental features
contaminating emission lines, prevented us from carrying out a reliable
classification.  Figure  \ref{diagnostic} shows that a large fraction of 
objects classified as star-forming galaxies lie close to the line delimiting
Seyfert from HII-galaxy regions of the parameter space.

Class (iii) consists of ten Seyfert 1 type objects, exhibiting broad
emission lines (Ly$\alpha$\,1216, [C\,IV]\,1549, [C\,III]\,1909,
[Mg\,II]\,2796) and nineteen Seyfert 2 type galaxies. 
Nine of the Seyfert 2 type sources will shift to the H\,II region of the 
parameter space in Figure \ref{diagnostic} if we correct their line ratios for
the mean H$\beta$ absorption of -1.7\AA\,  and the average dust
extinction $A_{V}$=1.0 (section 4.2)
These objects are designated with a question mark in Table 1.
Class (ii) includes 13 sources with absorption spectra
and a single (H$\alpha$) emission line.  These galaxies have a mean
redshift $z\approx0.1$ and are likely to be early-type spirals or
post-starburst galaxies.  Additionally, two objects exhibiting the
[Mg\,II]\,2796 absorption feature, characteristic of extreme local
starbursts (Guzman {\it et al.} 1997) are classified as star-forming
galaxies. Finally, 18 stars are found in the sample 
most of which lie at large 
angular separations from the radio position. The fraction of radio 
stars contributing to the radio counts at faint flux densities is difficult 
to estimate, since little is known about their radio luminosity functions.
Benn {\it et al.} (1993) found eleven stars within a sample of 87 
spectroscopically observed optical identifications with S$_{1.4}\ge0.2$\,mJy  
and $m_{V}\le20$ over an effective area of $\sim$1\,deg$^{2}$. They argue
that less than one radio star is expected in their sample and conclude that 
the rest of the stellar identifications are foreground contamination.
Because the present spectroscopic sample has similar radio flux density
and optical magnitude limits to that of Benn {\it et al.} (1993), we expect 
most of the stellar candidate identifications to be spurious.

\begin{figure} 
\centerline{\psfig{figure=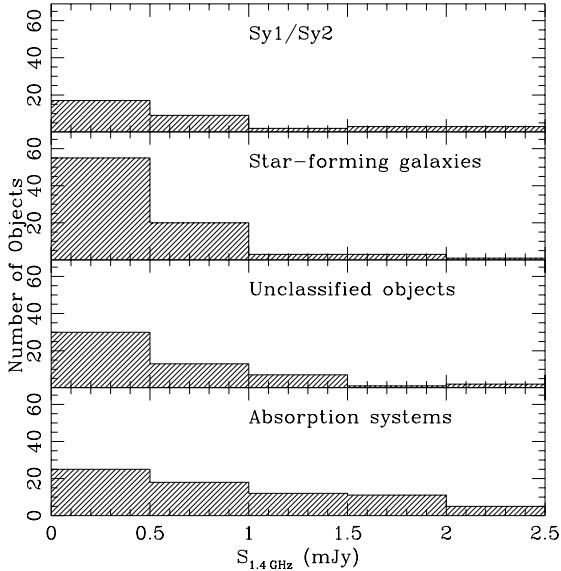,width=0.45\textwidth,angle=0}}
\caption{The distribution of 1.4 GHz flux density for the four optical
  spectroscopic classes.}\label{S14_hist}
\end{figure}

The radio flux density distribution as a function of the optical
classification is presented in Figure \ref{S14_hist}.  Star-forming 
galaxies dominate at sub-mJy levels (S$_{1.4}<0.5$\,mJy), with a small
contribution ($\approx10\%$) from Seyfert galaxies.  This agrees with
previous studies (Benn  {\it et al.} 1993; Kron {\it et al.} 1985).
Absorption line systems are the dominant population at flux densities $ >
$1\,mJy, although there is also a significant number at sub-mJy levels. 

\begin{figure} 
\centerline{\psfig{figure=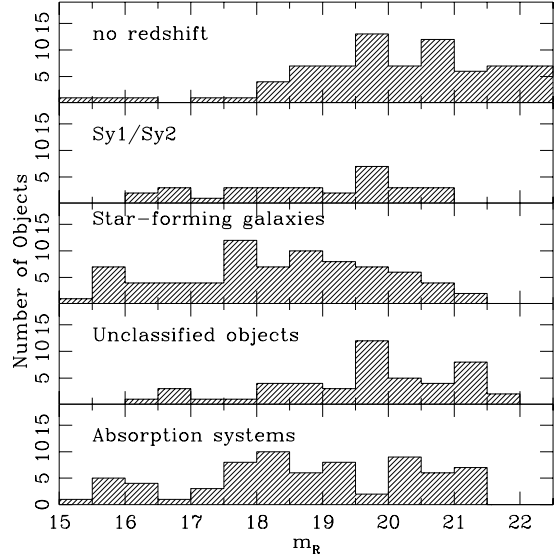,width=0.45\textwidth,angle=0}}
\caption{The distribution of apparent magnitude, $m_{R}$, for different classes
  of sources.}\label{mr_hist}
\end{figure}

The apparent magnitude distribution of the optical counterparts of faint
radio sources is presented in Figure \ref{mr_hist}.  There is some evidence
that for magnitudes m$_{R}>19.5$\,mag the proportion of star-forming galaxies
is decreasing relative to that of absorption line systems and
Seyferts. Gruppioni {\it et al.} (1998), using a faint (S$_{1.4}>$0.2\,mJy)
radio sample, combined with deep (m$_{R}<$22.0\,mag) spectroscopic data,
also found that the majority of the optical identifications fainter than
$m_{R}$=20.0\,mag are absorption line systems rather than star-forming
galaxies.  If this trend is real, it probably reflects differences in the
luminosity function of spirals and ellipticals. 

However, it has been demonstrated that evolutionary models of giant 
ellipticals and QSOs, that dominate the radio population 
at high flux densities, cannot easily reproduce the 
flattening in the normalised radio source counts at sub-mJy 
levels (Danese {\it et al.} 1987; Danese, De Zotti \& Franceschini 1985). 
Additionally, the interpretation of Figure 4 is difficult due to selection 
effects and  incompleteness at faint optical magnitudes. A significant 
fraction of the faintest objects have either unclassified spectra (see 
section 4.5), or no redshift determination. This may modify the distribution 
seen in Figure \ref{mr_hist}. Furthermore, at faint optical magnitudes the classification
scheme is less reliable due not only to the low S/N ratio of some of the 
spectra but also to the fact that fainter objects lie on average at higher 
redshift. As a result, not enough emission lines lie within the observable
window to perform a reliable classification ([O\,III]\,5007 redshifts outside
the observable window at $z\approx0.65$).

\begin{figure} 
\centerline{\psfig{figure=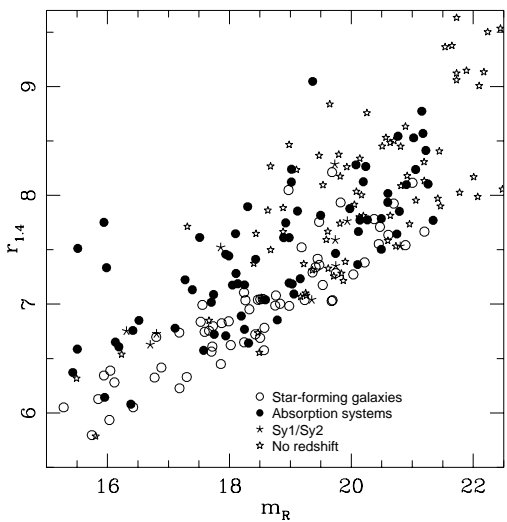,width=0.45\textwidth,angle=0}}
\caption{Radio-to-optical flux ratio versus apparent magnitude.}\label{r14_vs_mr}
\end{figure}

For optical identifications with no redshift determination some
information regarding their nature can be obtained using the
radio-to-optical flux ratio, $r_{1.4}$, defined as

\begin{equation}
r_{1.4}=\log\,S_{1.4}+0.4\,m_{R},
\end{equation}

\noindent where S$_{1.4}$, $m_{R}$ are the radio flux and the optical apparent 
magnitude of sources respectively. This ratio, although distance independent, 
is affected by the differential optical to radio K-correction. 
However, this is expected to be a second order effect. In Figure
\ref{r14_vs_mr}, $r_{1.4}$ is plotted  against m$_{R}$, for all the objects
detected spectroscopically. Star-forming galaxies, despite the scatter,
have on average lower radio-to-optical flux ratios  compared to absorption
line systems at a given magnitude. Furthermore, many of the sources
with no redshift determination, have $r_{1.4}$ values higher than those of
star-forming galaxies and similar to those of  ellipticals.  This indicates
that some of these sources, are likely to be  early type galaxies rather
than starbursts. 

\subsection{Visual extinction from H$\alpha$/H$\beta$}

To quantify the visual extinction, $A_{V}$, of the $z<0.3$ star-forming
galaxies that exhibit both H$\alpha$ and H$\beta$ emission lines,  
we compare the H$\alpha$/H$\beta$ decrement with the theoretical (case B
recombination) value of 2.86 (Brocklehurst 1971) and a standard reddening
curve (Savage \& Mathis 1979).  The technique is hampered by the poor S/N
ratio of some of the spectra, by the uncertainties in the flux calibration
(2dF data) and by stellar H$\beta$ absorption.

\begin{figure} 
\centerline{\psfig{figure=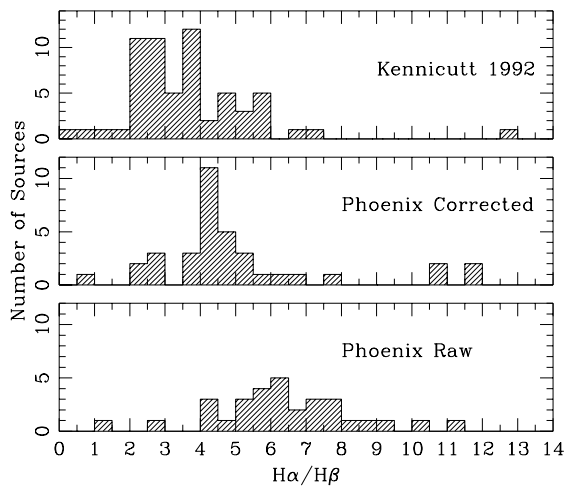,width=0.45\textwidth,angle=0}}
\caption{Distribution of the Balmer decrement for the optical
  counterparts of faint radio sources without correcting for H$\beta$
  absorption (bottom panel) and after accounting for that effect (middle
  panel).  The top panel is the distribution of H$\alpha$/H$\beta$
  for normal galaxies (Kennicutt 1992).  }
\end{figure}

Figure 6 (bottom panel) is a histogram of H$\alpha$/H$\beta$, not
corrected for stellar H$\beta$ absorption. The distribution peaks at a value of
6, significantly below the mean value of 12 found for IRAS galaxies (Leech
{\it et al} 1988; Elston, Cornell \& Lebofsky 1985).  Applying a correction
of $-1.7\pm0.3$\,\AA\, to H$\beta$ (Appendix $B$) lowers the peak of the
distribution to $\approx 4$ (Fig. 6, middle panel), consistent with
$A_{V}\approx1.0\pm0.1$ or $E(B-V)\approx 0.32\pm0.03$. 
The corrected distribution implies an average dust extinction similar to 
that found by Kennicutt (1992) for normal galaxies (Fig. 6, top panel) and 
comparable to the value $E(B-V)\approx 0.36$ obtained for disk H\,II regions 
(Kennicutt, Keel \& Blaha 1989). However, the dust reddening  derived here 
is smaller than that in nuclear H\,II regions ($E(B-V)\approx 0.56$; 
Kennicutt, Keel \& Blaha 1989), starburst galaxies ($E(B-V)\approx 0.69$; 
Dahari \& De Robertis 1988), IRAS sources ($E(B-V)\approx 1.3$; Leech 
{\it et al} 1988; Elston, Cornell  \& Lebofsky 1985) and H\,II luminous 
infrared galaxies ($E(B-V)\approx 0.99$; Veilleux {\it et al.} 1995).

It should be noted that  the objects selected for spectroscopic
observations are the optically brighter in the radio sample and therefore,
our spectroscopic sample is biased against very dusty sources.
Additionally, far-infrared and radio surveys have shown that the dust and
the molecular gas in luminous infrared galaxies are generally concentrated
into compact regions of diameter less than 1\,kpc (Scoville {\it et al.}
1986, 1989, 1991; Radford {\it et al.} 1991).  For the 2dF data, the fibre
diameter of 2\,arcsec corresponds to a physical size of 4 and  10\,kpc at
$z$=0.1 and 0.3 respectively. This implies that the 2dF spectra sample large
regions of the galaxy, which may possess a wide range of optical depths.
Depending on how the regions producing the detected line and continuum
emission are disposed relative to any sources of extinction, large
adjustments to the extinction derived here may be required, probably in the
upwards direction.

\subsection{Colour-redshift relations}

\begin{figure*} 
\centerline{\psfig{figure=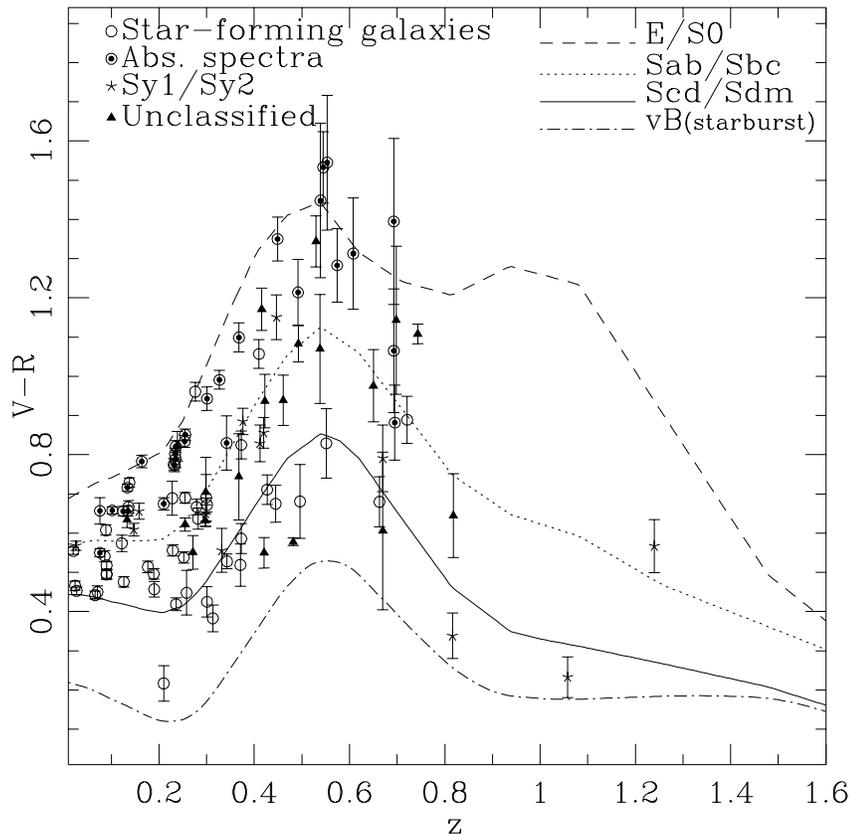,width=5in,height=5in,angle=0}} 
\caption{The $V-R$ colour-redshift diagram for the faint radio population,
  comprising star-forming galaxies ($\circ$), absorption systems ($\odot$),
  Sy1/Sy2 objects ($\star$) and unclassified objects ($\triangle$). The 
  curves are based on galaxy evolution models, as explained in the text:
  Dashed line - E/S0 model; Dotted line - Sab/Sbc model; Continuous line -
  Scd/Sdm model; Dash-dotted line - starburst (vB; very blue) galaxies.  }
\end{figure*}

Figure 7 compares the $V-R$ colour-redshift relation of the faint radio
population with the predictions of the evolutionary models of Bruzual \&
Charlot (1993; see Appendix $C$).  There is satisfactory agreement between
the observed photometric indices (i.e. colours) of the faint radio sources
and the completely independently classified types, based on their spectral
features and diagnostic emission line ratios.

Correcting the observed optical colours using $A_{V}=1.0$ (previous
section) will shift $V-R$ bluewards by $\approx$0.3 magnitudes (Savage \&
Mathis 1979), consistent with the late-type spiral and starburst (vB; very
blue) galaxy models.  Such blue optical colours would imply that
star-formation has not declined significantly in these galaxies since their
formation, in agreement with previous studies (Benn {\it et al.} 1993; Kron
{\it et al.} 1985).

\begin{figure*} 
\centerline{\psfig{figure=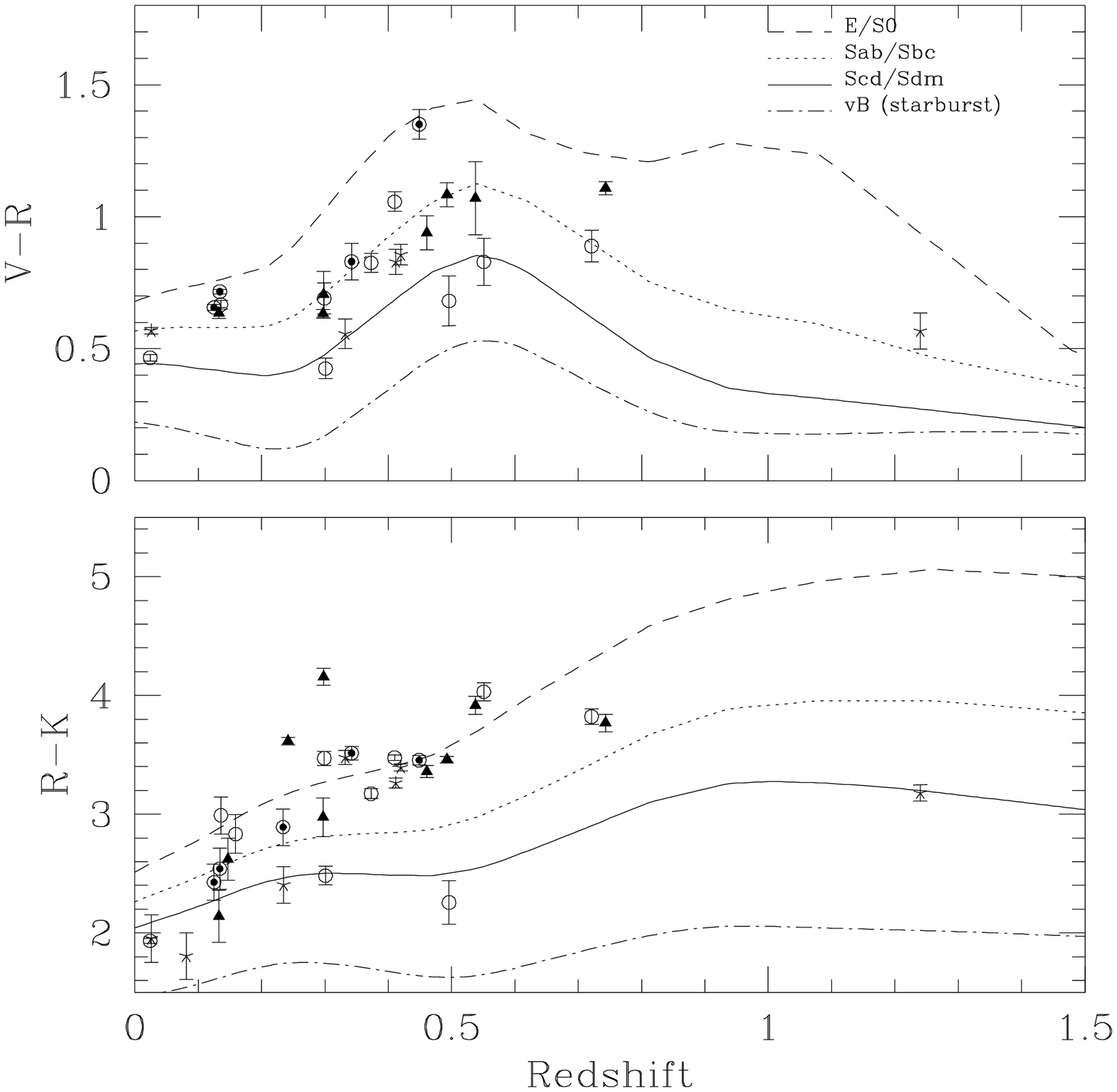,width=5in,height=5in,angle=0}} 
\caption{$V-R$ and  $R-K$  colour-redshift diagrams for radio 
  sources with near-infrared information. The symbols and the curves are
  defined in Figure 7.  }
\end{figure*}

\begin{figure*} 
\centerline{\psfig{figure=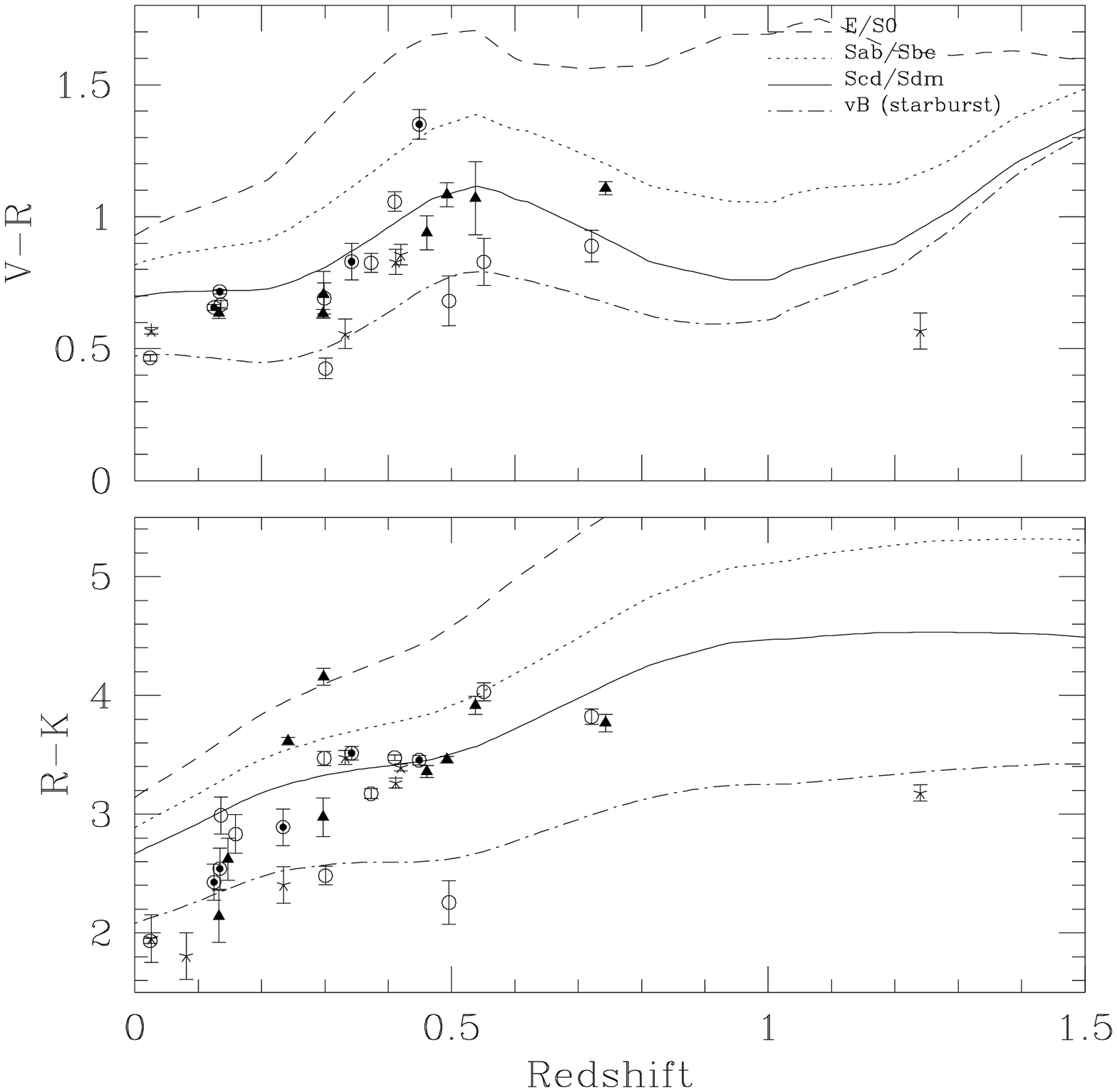,width=5in,height=5in,angle=0}} 
\caption{As for figure 8, with reddening ($A_{V}$=1.0) 
  included in the models.}
\end{figure*}

The $V-R$ and $R-K$ colour-redshift diagrams of the sub-mJy population with
$K$-band data are shown in Figure 8.  Near-infrared (2.2\,$\mu m$) emission
from galaxies is dominated by old stars, and is less affected by dust than
is optical emission.  Therefore, optical/near-infrared colours (e.g. $R-K$)
give information on the relative contributions from the giant and main
sequence populations.  Some galaxies in Figure 8 have $V-R$ bluer than the
early-type spiral (Sab/Sbc) model, but $R-K$ colours as red as those of
ellipticals.  Reddening due to extinction might help to explain this
apparent inconsistency. This possibility is illustrated in Figure 9 which
compares the $V-R$, $R-K$ colour-redshift diagram with models which also 
include dust reddening (Savage \& Mathis 1979) assuming  $A_{V}$=1.0.  A 
few galaxies still have inconsistent $V-R$ and $R-K$ colours which could 
be reconciled by adopting higher values of $A_{V}$.

Alternatively, the presence of an underlying evolved stellar population in
addition to main sequence stars could explain these observations.
Indications of two stellar populations (old giants and younger main
sequence stars) have been found out in brighter samples of radio galaxies
(`1\,Jy' sample; Allington-Smith 1982).  Lilly (1989) fit the observed
broad-band spectral energy distributions (SEDs) of high-redshift radio
sources using a combination of an old stellar component (age$<2
\times$10$^{9}$\,y) and a young population (age$<1\times$10$^{9}$\,y).
The young component contributes $\approx10\%$ to the rest frame emission at
5000\,\AA\, and an even smaller fraction of the mass.  Chambers and Charlot
(1990) modelled this behaviour with a single episode of star formation
lasting $<$10$^{8}$ years, plus a lower level of ongoing star formation.
However,
there are indications that the rest-frame UV and optical continuum of
powerful radio sources is contaminated by both non-thermal radiation from
the central engine and strong optical emission lines, undermining any
attempt to model their colours using simple stellar population synthesis
models (McCarthy 1993). Nevertheless, there is some evidence
that at least a fraction of the observed broad-band light of powerful radio
galaxies, especially at NIR wavelengths, has stellar origin (McCarthy 1993
and reference therein). In any case, the bright radio sources studied by
Lilly (1989) and Chambers \& Charlot (1990) are significantly different in
their nature from the fainter ones considered here and therefore, any
comparison of the properties of the two populations is not straightforward.

\subsection{Colour-Luminosity Relations}
In Figure 10 the ratio of radio to optical luminosity, $R_{1.4}$ is plotted 
against the intrinsic $V-R$ colour. $R_{1.4}$ is defined as

\begin{equation}
R_{1.4}=\log\,\mathrm{P_{1.4}}+0.4\,M_{R},
\end{equation}

\noindent where P$_{1.4}$ is the radio luminosity in W\,Hz$^{-1}$ and $M_{R}$
is the absolute $R$-band magnitude. Optical K-corrections are calculated 
using model SEDs from the population synthesis models of Bruzual \& Charlot 
(1993) (Appendix $C$; section 4.6).  The optical K-corrections for the Seyfert
 1, 2 type galaxies in this sample were estimated using a spectral index of 
$\alpha$=-0.5 ($S_{\nu}\propto \nu ^{\alpha}$). The radio luminosities were 
K-corrected assuming a spectral index $\alpha$=-0.8.

\begin{figure*} 
\centerline{\psfig{figure=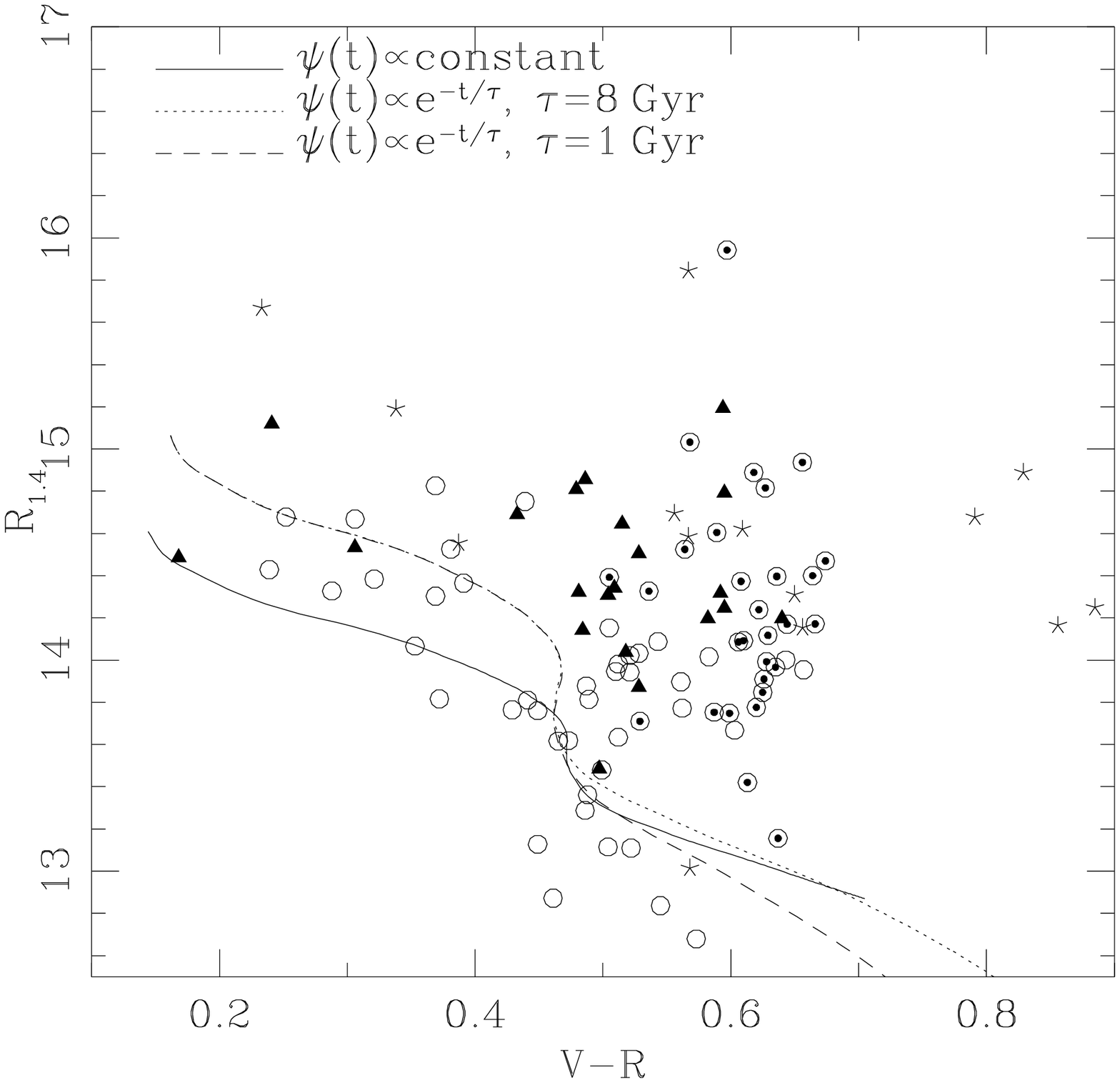,width=5in,height=5in,angle=0}} 
\caption{Radio-to-optical luminosity ratio against intrinsic $V-R$ colour.
  The curves are models with different star-formation rates, $\psi(t)$, as
  described in the text.  The symbols are the same as in Figure 7.}
\end{figure*}

In Figure 10 there are two populations of radio sources, (i) star-forming
galaxies and (ii) absorption line systems and Seyfert 1 and 2 type objects,
occupying different regions of the two-colour diagram. Star-forming
galaxies tend to have bluer colours for higher radio-to-optical emission. 
A similar result was obtained by Klein (1982) for a sample of galaxies
spanning a wide range of types.  The trend was interpreted as enhanced star
formation in bluer galaxies, producing massive stars, that (i) contribute
to the ionization of the interstellar gas and (ii) produce relativistic
electrons via supernova explosions, both leading to excess radio emission
(normalised to optical luminosity).  To demonstrate this trend,
we have constructed galaxy models with different parametric forms of the
star-formation rate, $\psi(t)$, as a function of time (Appendix $C$).  The
star-formation rate (SFR) at a given galaxy age is related to the radio
luminosity via (Condon 1992)

\begin{equation}
SFR(M\ge 5\,M_{\odot})=
\frac{\mathrm{P_{1.4}}}{4.0\times10^{21}\,W\,Hz^{-1}}\, M_{\odot}\,yr^{-1}.
\end{equation}
To compare with the observations, a reddening $A_{V}$=1.0 has been
included in the models.  The results shown in Figure 10 reproduce fairly
well the observed range of colours and radio-to-optical luminosity ratios
of star-forming galaxies.

A fraction of the unclassified sources in Figure 10 lie close to the region
of the parameter space occupied by ellipticals and Seyfert type
objects. These objects will be further discussed in the next section.

Absorption line systems exhibit a tight $V-R$ colour dispersion. This is
better demonstrated in Figure 11, presenting the colour-magnitude relation
for the optical counterparts of faint radio sources.  The small $V-R$
dispersion of the absorption line systems over a wide range of optical
luminosities agrees with the colour-magnitude relation (continuous line)
derived for normal ellipticals in the Virgo cluster (Sandage \& Visvanathan
1978). A similar result was obtained by Rixon {\it et al.} (1991), who 
also found a small dispersion ($<0.02$\,mag) in the rest-frame $V-R$ 
colours of radio selected early type galaxies with radio luminosities 
in the range $24.4<\log P_{408MHz}<26.3$\,W\,Hz$^{-1}$.  On the contrary, 
the star-forming  galaxies show significant scatter, a result that has 
also been found in other studies (Griersmith 1980; Schroeder 1996) and 
was attributed to (i) the presence of young stars and (ii) the intrinsic 
spread in the metallicity of galaxies.

\begin{figure*} 
\centerline{\psfig{figure=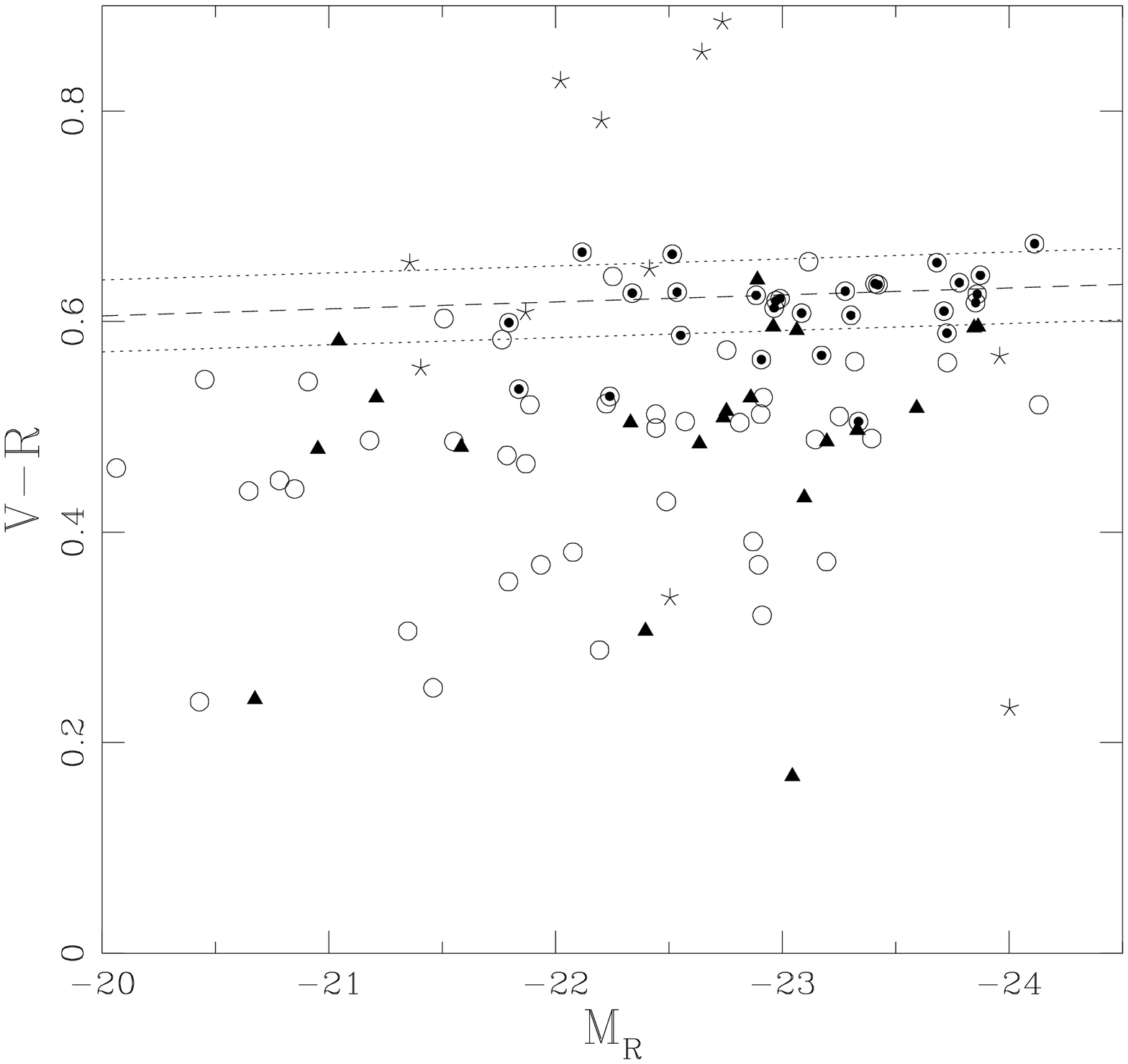,width=5in,height=5in,angle=0}}
\caption{Intrinsic $V-R$ colour against absolute $R$-band magnitude relation 
  of the optical counterparts of faint radio sources. The symbols are the same
  as in Figure 7.  Also shown are the C-M relation for ellipticals in the Virgo 
  cluster (dashed line) along with the 2$\sigma$ envelope lines (dotted
  lines).}
\end{figure*}

\subsection{Radio and Optical Luminosity Distributions} 

Figures 12 and 13 exhibit the distributions of the radio power and the
absolute optical magnitude.  Optically identified faint radio sources of
any type have, on average, optical luminosities in excess of M$^{*}_{R}$
(M$^{*}_{R}=-22.5$\,mag; Metcalfe {\it et al.} 1991). This agrees with
other studies of radio galaxies at sub-mJy flux density levels (Benn {\it
  et al.} 1993; Hammer {\it et al.} 1995; Windhorst {\it et al.} 1995).
The absorption line systems (i.e. ellipticals) have optical magnitudes and
radio powers in the range -22.0$<M_{R}<$-24.5\,mag and
22.0$<\log\,\mathrm{P_{1.4}}<$25.0\,W\,Hz$^{-1}$ respectively, consistent 
with them being Fanaroff-Riley I types objects (Ledlow \& Owen 1996).
Furthermore, the distribution of these sources is tighter and peaks at a
brighter luminosity both at radio (by $\approx$1\,dex) and optical (by
$\approx$0.5\,mag) wavelengths, compared to star-forming galaxies.  This
latter class of objects spans the range -19.0$<M_{R}<$-24.0\,mag at optical
and 20.5$<\log\,\mathrm{P_{1.4}}<$24.5\,W\,Hz$^{-1}$ at radio wavelengths.

\begin{figure} 
\centerline{\psfig{figure=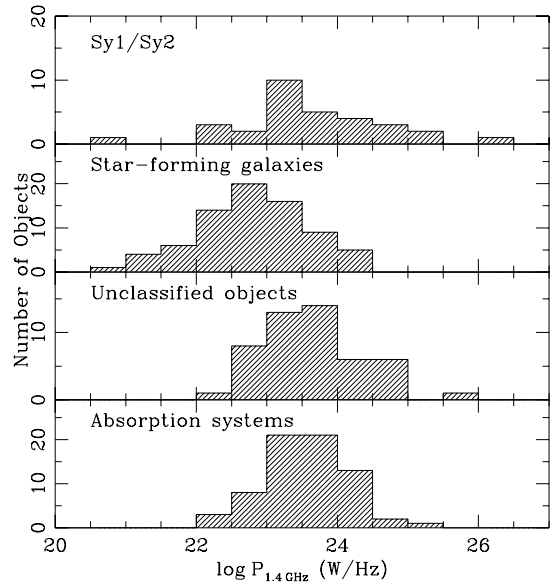,width=0.45\textwidth,angle=0}}
\caption{Distribution of radio powers for different classes of objects.}
\end{figure}

\begin{figure} 
  \centerline{\psfig{figure=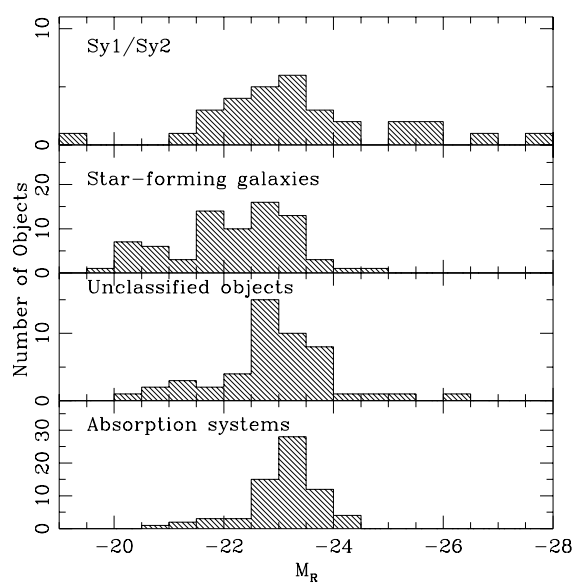,width=0.45\textwidth,angle=0}}
\caption{Distribution of optical luminosity for different classes of 
  objects.  }
\end{figure}

In Figure 12, a number of radio sources with unclassified spectra 
and radio power $>$23.5\,W\,Hz$^{-1}$ are brighter at radio wavelengths 
than star-forming galaxies (their distribution peaks at
$\log\mathrm{P_{1.4}}\approx23.0\,\mathrm{W\,Hz^{-1}}$) and as bright as
ellipticals or Seyfert type objects. Their median rest-frame $V-R$ colour
and radio-to-optical luminosity  ratio are $\approx$0.5 and $\approx$14.25 
respectively.  They are the unclassified sources lying close to the
elliptical/Seyfert region of the parameter space in Figure 10.  The median
redshift of these sources is $z \approx 0.5$, determined from only one
emission line (and in some cases, one or more absorption lines), which was
identified as [O\,II]\,3727 with rest frame equivalent width of the order
of $\approx 15$\,\AA.  The evidence above indicates that a fraction of the
high-$z$ unclassified sources may differ from star-forming galaxies.  They
might be elliptical galaxies powered by an AGN and exhibiting weak
[O\,II]\,3727 emission line, or spirals harbouring a central non-thermal
source in addition to any star-formation activity.

These objects are also relatively faint (m$_{R}\approx$20.0\,mag), implying
that a fraction of the optically faint unclassified sources in 
Figure 4 (section 4.1) might be ellipticals or Seyferts rather than 
starbursts. However, the evidence above remains circumstantial and further  
spectroscopic observations are essential to investigate the nature of the
central ionising source of these objects.

\subsection{Radio versus H$\alpha$ Luminosity}

Study of the relation between L$_{\mathrm{H\alpha}}$ and P$_{1.4}$ luminosities
would help elucidate the link between active star-formation and radio
emission in these galaxies. A relation between these two quantities for
star-forming galaxies has already been established (Benn {\it et al.}
1993), suggesting that the radio luminosity is a probe of on-going
star-formation in some of the sub-mJy sources.

In the absence of absolute flux calibration for the 2dF spectroscopic
observations, the H$\alpha$ luminosity is calculated using the $R$-band
magnitudes and the measured H$\alpha$+[N\,II] equivalent widths
(EW$_{\mathrm{H\alpha}+[NII]}$) of radio sources.  We employ the same method to
also calculate the H$\alpha$ luminosity for the objects observed
spectroscopically at ESO 3.6\,m telescope.  The procedure followed is
summarised below.

Firstly, the absolute $R$-band magnitude, M$_{R}$, of a given radio galaxy
is calculated from the relation

\begin{equation}
  M_{R}=m_{R}-5\times \log(d_{L}/10)-k_{R}(z),
\end{equation}

\noindent where $d_{L}$ is the luminosity distance and $k_{R}(z)$ 
is the K-correction
(Yoshii \& Takahara 1988). This latter is calculated using model SEDs of
different galaxy types from the population synthesis code of Bruzual
\& Charlot (1993; Appendix $C$).  The radio sources are assigned a model
galaxy type (E/S0, Sab/Sbc, Scd/Sdm or vB), based on their position on the
$V-R$ versus redshift diagram and the corresponding K-correction is then
calculated.  For those objects for which only $R$-band data are available,
we apply the average K-correction for the Sab/Sbc SED (Appendix $C$).  The
absolute magnitude is then converted to luminosity (L$_{R}$) using the
luminosity of Vega in the $R$-band
(1.869\,10$^{-12}$\,W\,m$^{-2}$\,\AA$^{-1}$) and the relation

\begin{equation}
 \mathrm{L_{R}=L_{Vega}}\times 10^{(-0.4 \times M_{R})}.
\end{equation}

\noindent Finally, having established L$_{R}$ and the model SED 
corresponding to a given radio source, we estimate, the continuum flux 
at 6563\,\AA\, ($C_{6563}$) from the model.  L$_{\mathrm{H\alpha}}$ is subsequently
estimated from  the relation

\begin{equation}
 \mathrm{L_{Ha}}=\frac{3}{4}\,C_{6563}\,EW_{\mathrm{H\alpha}+[N\,II]},
\end{equation}

\noindent where the factor 3/4 converts the H$\alpha$+[N\,II] equivalent 
width to that of H$\alpha$ (Kennicutt 1992). In the case of high redshift
galaxies ($z>0.3$), where H$\alpha$ is redshifted outside the observable window, 
we use [O\,II]\,3727 emission line to measure the star formation, relating 
the [O\,II]\,3727 equivalent width to that of H$\alpha$ via

\begin{equation}
 EW_{[\mathrm{OII}]}=0.4\times EW_{\mathrm{H\alpha}+[N\,II]},
\end{equation}

\noindent derived by Kennicutt (1992) for a sample of local galaxies 
spanning a wide range of types.

\begin{figure*} 
\centerline{\psfig{figure=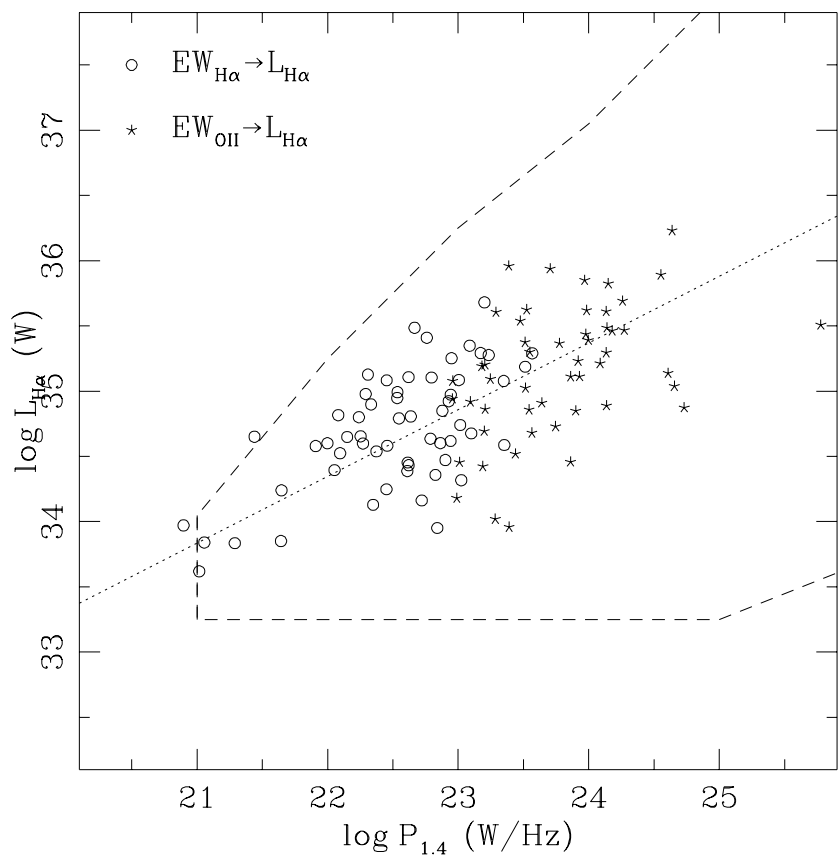,width=5in,height=5in,angle=0}}
\caption{Relation between radio power and  H$\alpha$ luminosity. Open
 circles ($\circ$)  correspond to radio galaxies with $z<0.3$,  where
 H$\alpha$ lies within the observable window. Asterisks ($\star$)
 correspond to radio galaxies at higher redshifts, where H$\alpha$
 is  redshifted outside the optical window.  For these objects we employ
 [O\,II]\,3727 as a proxy to H$\alpha$.  The dotted line represents the best
 fit to the data L$_{\mathrm{H\alpha}}\propto$P$_{1.4}^{-0.6}$. The dashed line
 delimits the region of the parameter space affected by the selection
 effects, as described in the text.}
\end{figure*}

The L$_{\mathrm{H\alpha}}$ versus P$_{1.4}$ relation is shown in Figure 14. To
remove any artificial trend in this relation due to incompleteness, we only
consider galaxies with optical magnitudes brighter than m$_{R}=21.5$\,mag 
and radio fluxes brighter than 0.4\,mJy, in the case of the PDF, or brighter 
than 0.2\,mJy in the case of PDFS.  Assuming a linear relation between 
P$_{1.4}$ and L$_{\mathrm{H\alpha}}$ we find L$_{\mathrm{H\alpha}}=10^{12}$\,P$_{1.4}$, in 
excellent agreement with that independently found by Benn {\it et al.} (1993).
However, the best fit to the data is a power law of the form
L$_{\mathrm{H\alpha}}\propto$P$_{1.4}^{-0.6}$ (dotted line in Figure 14).  This
relation also holds if the $z<0.3$ and the $z>0.3$ sources are treated
separately.

However, Figure 14 is not free of selection effects.  Firstly, the radio
survey is peak flux density limited and complete to $\approx$0.4\,mJy.
Secondly, the spectroscopic observations are carried out only for the
optically identified galaxies in the radio sample, introducing a bias
against optically faint sources.

To confirm the reality of the trend seen in Figure 14, we will assume that
L$_{\mathrm{H\alpha}}$ and P$_{1.4}$ are uncorrelated and we will investigate if
the selection effects introduce any artificial relation between these two
quantities.  We adopt a limiting optical magnitude $m_{R}$=21.5\,mag,
corresponding to the faintest object for which a redshift has been
estimated and a radio flux density limit of 0.4\,mJy.
To calculate the optical K-corrections we adopt the Sab/Sbc galaxy model
generated by the population synthesis code of Bruzual and Charlot (1993;
Appendix $C$). This will also be used to calculate L$_{\mathrm{H\alpha}}$ via
(Kennicutt 1992)

\begin{equation}
SFR=\frac{\mathrm{L_{H\alpha}}}{8.9\times 10^{34}\,W}\,M_{\odot}\,yr^{-1},
\end{equation}
where the star formation rate (SFR) is given by the model.  An average
correction of 1.0\,mag, to allow for dust extinction (Kennicutt 1992), is
also applied to this equation.

The apparent magnitude m$_{R}$ and flux density S$_{1.4}$, corresponding to
a given set of values M$_{R}$ (M$_{R}<$-20.0\,mag) and P$_{1.4}$
(P$_{1.4}>$20.0\,W\,Hz$^{-1}$) respectively, are calculated, for redshifts
in the range $0<z<1.5$ (corresponding to the $z$ range of the sources in
our sample). If m$_{R}$ and S$_{1.4}$ lie within the optical and radio 
completeness limits respectively, the values of M$_{R}$ and P$_{1.4}$ are
registered.  Following this procedure, we define the bias free region in
Figure 14, demarcated by the dashed lines.  It is obvious that the sample
selection is not significantly affecting our results.

Furthermore, the 2dF fibre size might also introduce a bias into our
sample. If the H$\alpha$ emitting region is large, we may have
underestimated the H$\alpha$ equivalent width and hence the H$\alpha$
luminosity, for nearby extended galaxies. That could induce a false
correlation in Figure 14 (Leech {\it et al.} 1988).  However, most of the
sources in our sample, lie at redshifts $>0.1$ and therefore this effect is
expected to be small.

We conclude that the strong correlation between radio power and
H$\alpha$ luminosity in Figure 14 is real. This is interpreted as the
result of the tight correlation between far-infrared and radio power
(Condon 1992) and H$\alpha$ and far-infrared luminosities (Leech {\it et
  al} 1988). Therefore, the radio power can also be used as a probe of the
star-formation activity in starburst galaxies.  Additionally, the
insensitivity of radio luminosities to dust extinction compared to
H$\alpha$ provides further  advantage in using this to study
 the star-formation.

\section{Conclusions}
The 1.4 GHz Phoenix radio survey has provided a homogeneous sample of
sources which we have used to study the nature of the faint radio
population. Our conclusions are summarised below:

\begin{enumerate}
\item many of the optical counterparts of sub-mJy radio sources have $V-R$
  colours and spectroscopic properties, similar to those of star-forming
  galaxies. However, radio sources with absorption line spectra, similar to
  those of ellipticals, are also found at sub-mJy levels.
\item the average visual extinction of the $z<0.3$ star-forming galaxies
  that host faint radio sources appears to be similar to that of normal
  galaxies (Kennicutt 1992).
\item the $V-R$ and $R-K$ colours of the optical counterparts of faint
  radio galaxies are generally consistent with moderate amounts of
  extinction, with a small fraction of galaxies having evidence of higher
  extinction.
\item simple population synthesis models with dust reddening can
  reproduce the observed range of radio-to-optical luminosity ratios 
  and the rest-frame $V-R$ colours of star-forming radio galaxies.
\item the radio power of star-forming galaxies correlates with H$\alpha$
  luminosity, indicating that the former probes the active star-formation
  in these objects.
\end{enumerate}

\section{Acknowledgements} 
We wish to thank an anonymous referee for several constructive comments. 
AG is supported by a scholarship from the State Scholarships Foundation of 
Greece (S.S.F.). The work of LC and AH is supported by the Australian
Research Council and the Science Foundation for Physics within the
University of Sydney.

\appendix 
\section{Star-Galaxy separation}
The galaxy number counts provide a useful consistency check between our
optical catalogue and other surveys.  However, although the FOCAS number
counts at faint magnitudes are dominated by galaxies, at bright magnitudes
($m_{R}<$19.0, Hudon \& Lilly 1996), the stellar component makes a
significant contribution. To correct for this stellar contamination we
adopt the following procedure.  We considered a size parameter, defined as
the difference between the `core', $m^{R}_{core}$, and `total',
$m^{R}_{total}$, magnitude of sources.  The former corresponds to the
intensity within the 3$\times$3 inner pixels of an object and the latter is
the integrated luminosity within an aperture defined by adding rings around
the object until the detection area is exceeded by a factor of 2.  This
difference is then plotted against $m^{R}_{total}$ (Fig. A1).  Since the
size parameter is sensitive to the seeing, the data taken in different
conditions, should be treated separately.  Consequently, we grouped the
images of a given night according to the seeing conditions at the time of
the observation and performed the star-galaxy separation for each group
individually.  This is demonstrated in Figure A1 where the size parameter
is plotted against the `total' magnitude for frames with similar seeing
conditions, with the stellar sequence demarcated with a solid lined
rectangular box.  The distribution of stars and galaxies in this figure
overlaps at magnitudes fainter than $m_{R}$=20.0\,mag.  Beyond that
magnitude limit no attempt is made to further eliminate stars from the
sample, since compact galaxies could be mistakenly removed. Furthermore,
the number of stars relative to galaxies becomes increasingly smaller
beyond this magnitude.

\begin{figure} 
\centerline{\psfig{figure=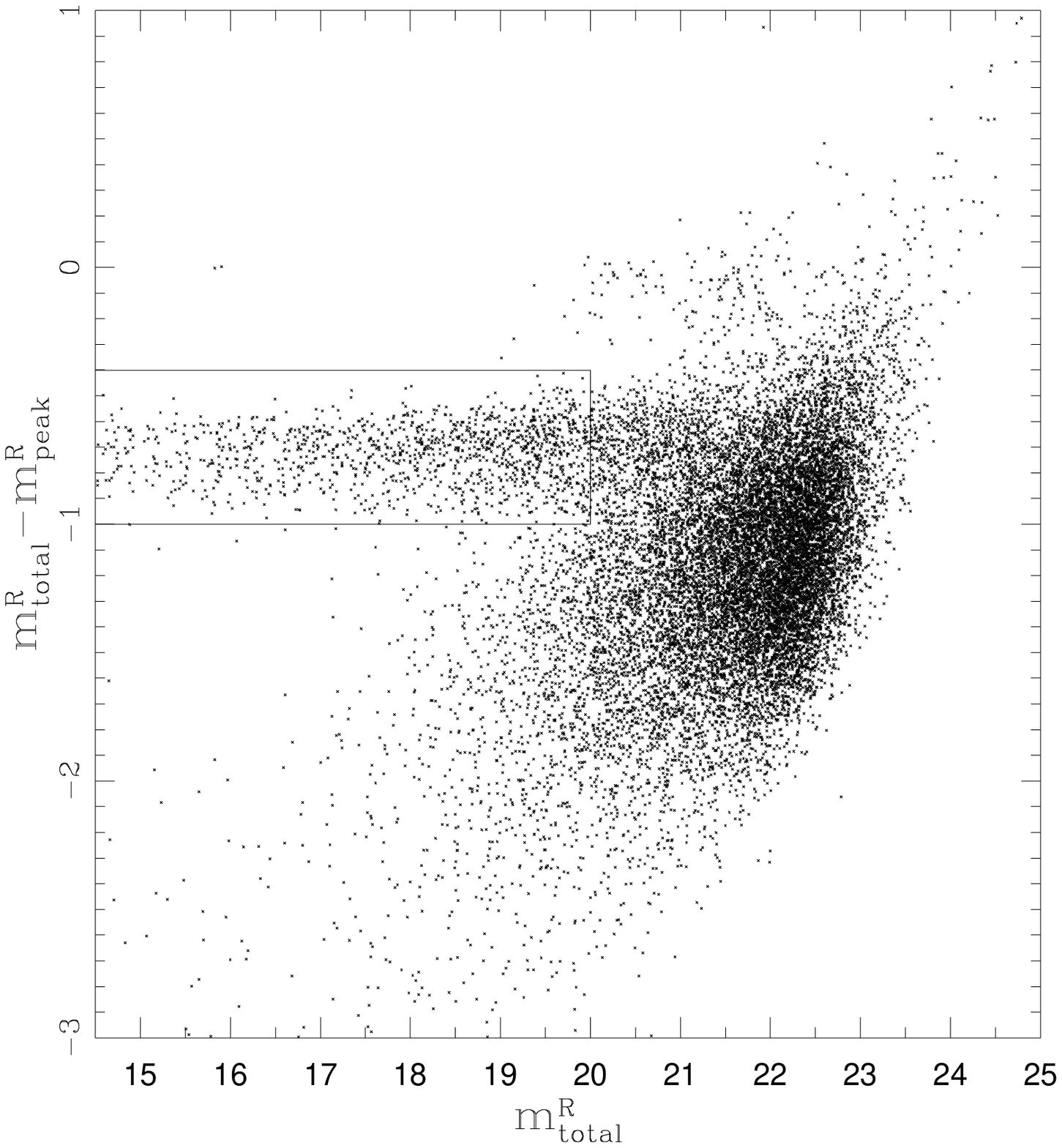,width=0.45\textwidth,angle=0}} 
\caption {Star-galaxy separation diagram for a subset of the  
  $R$-band observations performed in similar seeing conditions.  The
  stellar sequence is demarcated by a box.}
\end{figure}

The galaxy number counts are shown in Figure A2 along with the compilation
of Metcalfe {\it et al.} (1991).  For $m_{R}<22.5$\,mag our counts seem to
be in good agreement with the recent determinations.  At fainter magnitudes
the sample is affected by incompleteness.  Additionally, the star counts in
Figure A2 (small circles) are also in good agreement with the predictions
of the model of Bahcall \& Soneira (1980).

\begin{figure} 
\centerline{\psfig{figure=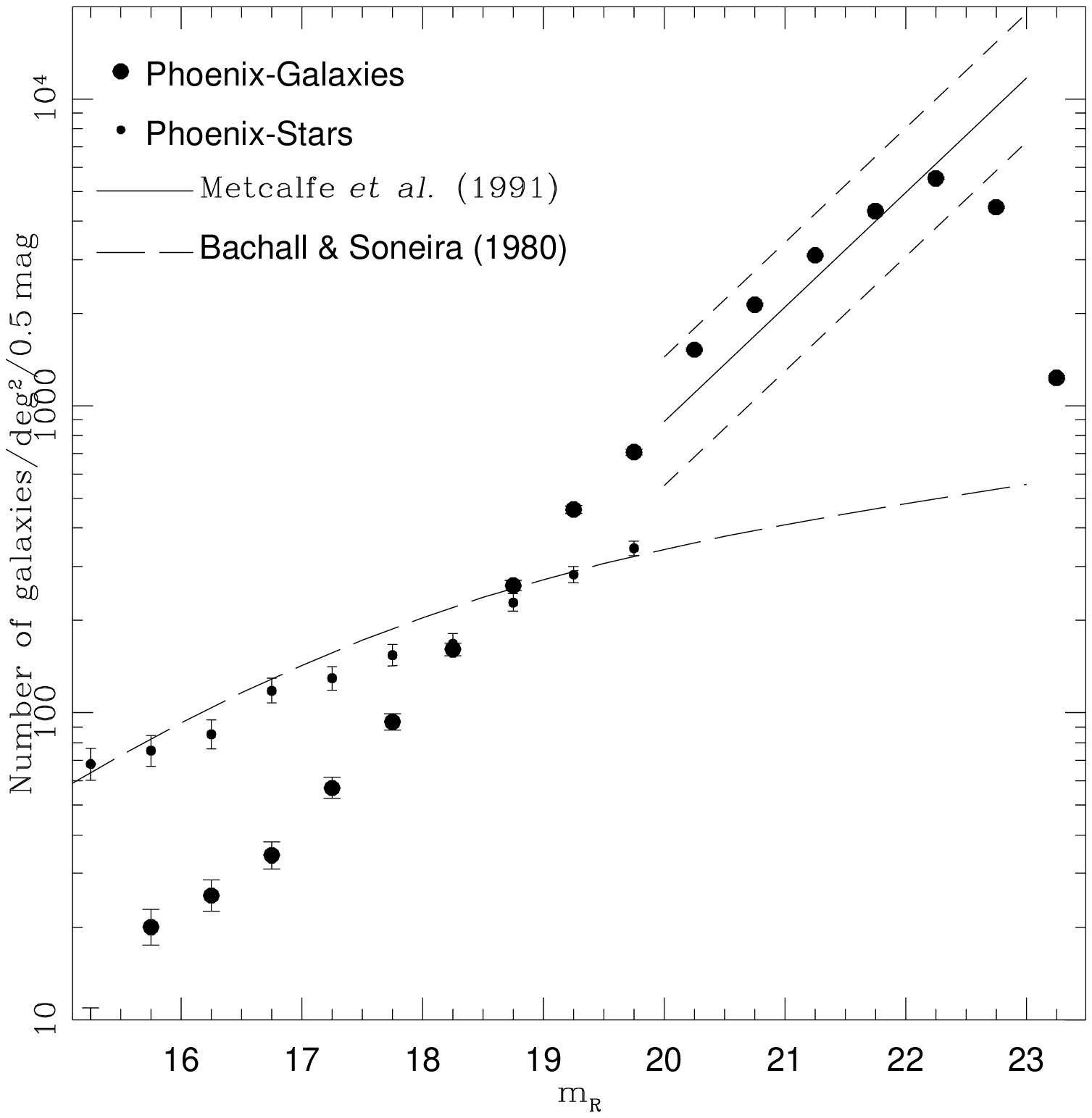,width=0.45\textwidth,angle=0}}
\caption{$R$-band galaxy counts (big circles) of
  the Phoenix field. The continuous line is the best fit to the galaxy
  source counts in the range 20$<m_{R}<$23 by Metcalfe {\it et al.}
  (1991). The dashed lines are the 1$\sigma$ envelope lines. Also shown are
  the star counts (small circles) from the present study, along with the
  predictions of the model of Bahcall \& Soneira (1980).}
\end{figure}

\section{H$\beta$ absorption}

To estimate the average H$\beta$ absorption for the sample of $z<0.3$
radio sources (where H$\alpha$ lies within the observable window),
we consider the relation between the H$\beta$ and H$\alpha$+[N\,II]
equivalent widths for star-forming galaxies and Seyfert 2 type objects, 
presented in Figure B1. Additionally, radio galaxies with 
absorption line spectra (for which EW$_{\mathrm{H\beta}}\le0$) and only 
H$\alpha$  emission 
line, classified star-forming galaxies in  section 4.1, 
are also plotted in Figure B1. The best fit to the data, shown with the 
continuous line in that Figure, follows the relation

\begin{equation}\label{eq_B1}
 EW_{\mathrm{H\beta}}=EW_{\mathrm{H\alpha+[N\,II]}}/(9.1\pm0.4)-(1.7\pm0.3).
\end{equation}

\noindent The coefficients in equation (B1) imply a mean H$\beta$
absorption of $-1.7\pm0.3$\,\AA.  Kennicutt (1992), following  a 
similar procedure, found an  average stellar H$\beta$ absorption  of 
$\approx -5$\,\AA\, for normal galaxies. However, a value of -2\,\AA\, 
is usually used in the literature for H\,II regions 
(Tresse {\it et al.} 1996), in good agreement with our result.

\begin{figure}
\centerline{\psfig{figure=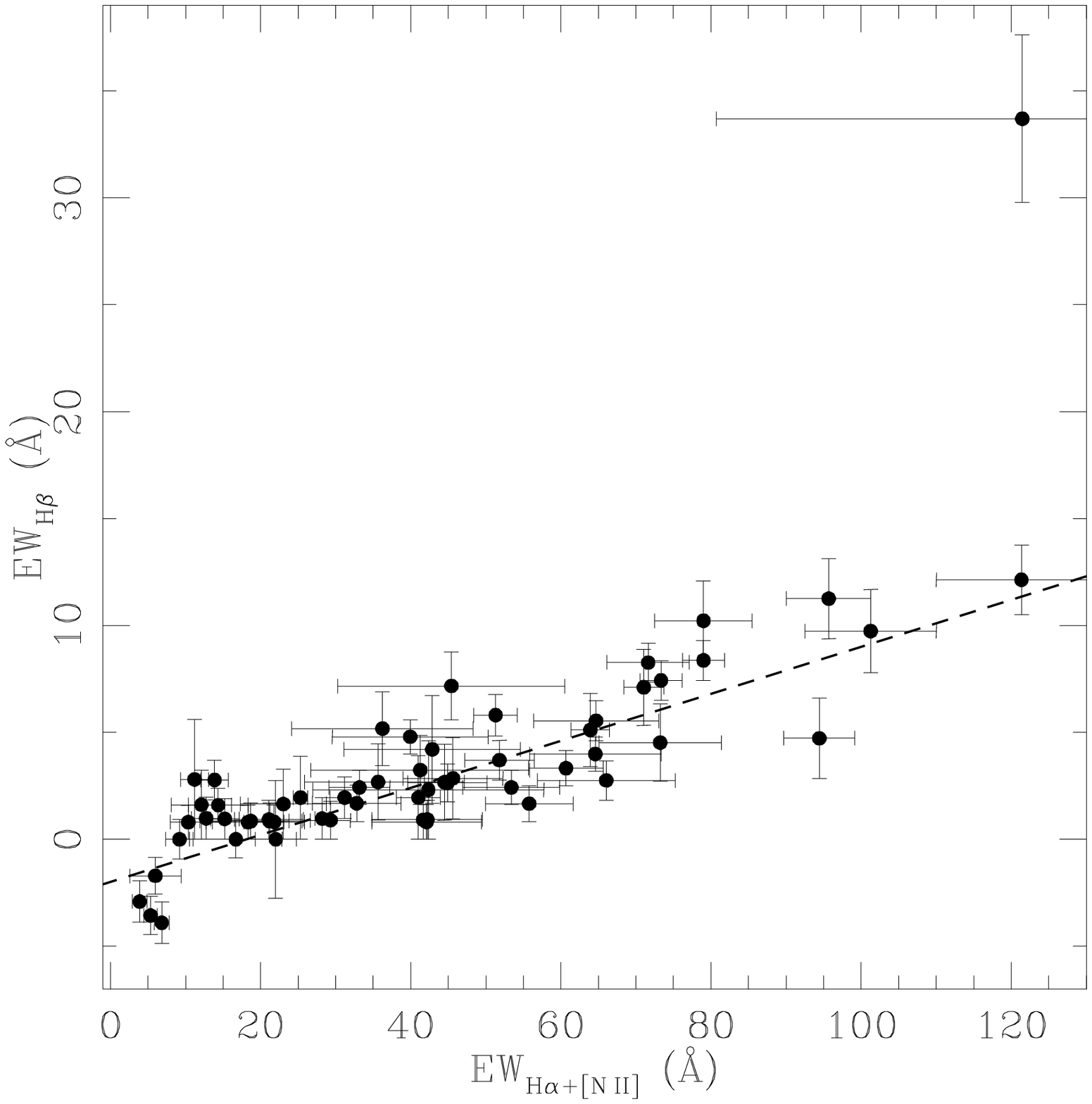,width=4in,height=4in,angle=0}}
 \caption{Relation between the  H${\beta}$ and H${\alpha}$+[N\,II]
 equivalent widths. The dashed line is the best fit to the data, given by
 the relation EW$_{\mathrm{H\beta}}$=EW$_{\mathrm{H\alpha+[N\,II]}}/(9.1\pm0.4)-(1.7\pm0.3)$.}
\end{figure}

\section{Population Synthesis Models}
In this study we adopt the galaxy models developed by Pozzetti, Bruzual \&
Zamorani (1996) that make use of the population synthesis code of Bruzual
\& Charlot (1993) to generate evolutionary models for ellipticals (E/S0),
early (Sab/Sbc) and late (Scd/Sdm) type spirals and very blue galaxies
(vB), meant to reproduce the starburst population present at each redshift.

Assuming $H_{o}=50$\,km$^{-1}$\,s$^{-1}$\,Mpc$^{-1}$, $q_{o}=0.5$ and a
formation redshift of $z_{f}=10$, corresponding to an age of 12.7\,Gyr, the
adopted parameters for the different galaxy types, are listed in Table C1.
The local properties of E/S0 and Sab/Sbc galaxies are best represented by
an exponentially decaying Star Formation Rate (SFR), of the form
$\psi(t)\propto\tau^{-1}exp(-t/\tau)$, where $\tau$ is the e-folding time.
A value of $\tau$=1 and 8\,Gyr was adopted for the E/S0 and Sab/Sbc
galaxies respectively.  Late type spirals and starbursts are best
represented by a constant SFR and different ages.

\begin{table} 
\footnotesize 
\begin{center} 
\begin{tabular}{cccc} 
 
 Model SED & SFR & IMF & Age (Gyr)\\  
 E/S0     &   $\tau_{1}$     &  Scalo          & 12.7 \\  
 Sab/Sbc  &   $\tau_{8}$     &  Scallo         & 12.7 \\  
 Sbc/Scd  &   constant       &  Salpeter       & 12.7 \\  
  vB      &   constant       &  Salpeter       & 0.1 \\  
\end{tabular} 
\end{center} 
\caption{The Star-Formation Rate (SFR), Initial Mass Function (IMF)
  and present day galaxy age adopted to construct the Spectral Energy
  Distributions (SEDs) of different galaxy types.} \normalsize 
\end{table} 

\end{document}